\documentclass[trackchanges, twocolumn]{aastex701}

\usepackage{float}
\usepackage{amsmath}
\usepackage{gensymb}
\usepackage{tikz}
\usetikzlibrary{decorations.pathreplacing,calligraphy}

\begin{document}

\title{Using Machine Learning to Model Stellar Collisions in our Galactic Center}

\author[orcid=0009-0004-5934-9650,gname='Tristan', sname='Sand']{Tristan C. Sand}
\affiliation{Department of Physics, Loyola University Chicago, Chicago, IL 60660, USA}
\affil{Department of Physics \& Astronomy, Northwestern University, Evanston, IL 60208, USA}
\affil{Center for Interdisciplinary Exploration \& Research in Astrophysics (CIERA), Northwestern University, Evanston, IL 60201, USA}
\email[show]{tristan.sand@northwestern.edu}  

\author[orcid=0000-0003-0984-4456,gname='Sanaea', sname='Rose']{Sanaea C. Rose} 
\affil{Center for Interdisciplinary Exploration \& Research in Astrophysics (CIERA), Northwestern University, Evanston, IL 60201, USA}
\affiliation{NSF-Simons AI Institute for the Sky (SkAI), 172 E. Chestnut St., Chicago, IL 60611, USA}
\email{sanaea.rose@northwestern.edu}

\author[orcid=0000-0002-0933-6438, gname='Elena', sname='González Prieto']{Elena Gonz\'{a}lez Prieto}
\affil{Department of Physics \& Astronomy, Northwestern University, Evanston, IL 60208, USA}
\affil{Center for Interdisciplinary Exploration \& Research in Astrophysics (CIERA), Northwestern University, Evanston, IL 60201, USA}
\affil{NSF-Simons AI Institute for the Sky (SkAI), 172 E. Chestnut St., Chicago, IL 60611, USA}
\email{elena.prieto@northwestern.edu}

\author[orcid=0000-0002-7444-7599, gname='James',sname='Lombardi']{James C. Lombardi, Jr.}
\affiliation{Department of Physics, Allegheny College, Meadville, Pennsylvania 16335, USA}
\email{jalombar@allegheny.edu}

\author[orcid=0009-0003-8690-8297, gname='Charles',sname='Gibson']{Charles F. A. Gibson}
\affil{Department of Physics \& Astronomy, Northwestern University, Evanston, IL 60208, USA}
\affil{Center for Interdisciplinary Exploration \& Research in Astrophysics (CIERA), Northwestern University, Evanston, IL 60201, USA}
\affiliation{NSF-Simons AI Institute for the Sky (SkAI), 172 E. Chestnut St., Chicago, IL 60611, USA}
\email{charlesgibson2031@u.northwestern.edu}

\author[orcid=0000-0003-4412-2176, gname='Fulya',sname='Kıroğlu']{Fulya Kıroğlu}
\affil{Center for Interdisciplinary Exploration \& Research in Astrophysics (CIERA), Northwestern University, Evanston, IL 60201, USA}
\affiliation{NSF-Simons AI Institute for the Sky (SkAI), 172 E. Chestnut St., Chicago, IL 60611, USA}
\email{FulyaKiroglu2024@u.northwestern.edu}

\author[orcid=0000-0002-4086-3180, gname='Kyle',sname='Kremer']{Kyle Kremer}
\affiliation{Department of Astronomy \& Astrophysics, University of California, San Diego; La Jolla, CA 92093, USA}
\email{kykremer@UCSD.edu}

\author[orcid=0000-0003-3987-3776, gname='Christopher',sname='O'Connor']{Christopher E. O'Connor}
\affil{Center for Interdisciplinary Exploration \& Research in Astrophysics (CIERA), Northwestern University, Evanston, IL 60201, USA}
\affiliation{NSF-Simons AI Institute for the Sky (SkAI), 172 E. Chestnut St., Chicago, IL 60611, USA}
\email{christopher.oconnor@northwestern.edu}

\author[0000-0003-2558-3102]{Enrico Ramirez-Ruiz}
\affiliation{Department of Astronomy and Astrophysics, University of California, Santa Cruz, CA 95064, USA}
\email{enrico@ucolick.org}

\author[orcid=0000-0002-7132-418X, gname='Frederic', sname='Rasio']{Frederic A. Rasio}
\affil{Department of Physics \& Astronomy, Northwestern University, Evanston, IL 60208, USA}
\affil{Center for Interdisciplinary Exploration \& Research in Astrophysics (CIERA), Northwestern University, Evanston, IL 60201, USA}
\affiliation{NSF-Simons AI Institute for the Sky (SkAI), 172 E. Chestnut St., Chicago, IL 60611, USA}
\email{rasio@northwestern.edu}

\author[orcid=0000-0003-2539-8206, gname='Tjitske',sname='Starkenburg']{Tjitske Starkenburg}
\affil{Department of Physics \& Astronomy, Northwestern University, Evanston, IL 60208, USA}
\affil{Center for Interdisciplinary Exploration \& Research in Astrophysics (CIERA), Northwestern University, Evanston, IL 60201, USA}
\affiliation{NSF-Simons AI Institute for the Sky (SkAI), 172 E. Chestnut St., Chicago, IL 60611, USA}
\email{tjitske.starkenburg@northwestern.edu}

\begin{abstract} \label{abstract}

Direct collisions in the inner pc of the Galactic center can alter the orbits and properties of stars. The outcome of a collision depends on a number of parameters, including the masses and ages of the stars, the impact parameter, and the initial relative velocity. 
We utilize the newly developed \verb|collAIder|, a machine learning tool developed to predict the outcomes of stellar collisions, to bridge 3D hydrodynamic simulations of stellar collisions with a dynamical model of the Galactic center. 
Our results show a substantial population of stripped stars, $\sim$$10\%$ of the initial stellar population, 
produced by mass loss during high-speed collisions. Stellar mergers are less frequent, with about $5\%$ of the stars experiencing a collision-induced merger. We also find that high-speed, nearly head-on collisions can completely disrupt the stars. These destructive collisions are usually preceded by 10 or more collisions, which gradually reduce the mass of the star before it is ultimately destroyed. We estimate that mass loss during stellar collisions injects roughly $(2.5-8)\times 10^5$~M$_\odot$ of gas into the surrounding environment. Most of this gas is injected into the inner $0.1$ pc and about $90\%$ is retained within the cluster and may be accreted onto the supermassive black hole. 

\end{abstract}


\section{Introduction} \label{sec:intro}

Galaxies similar to the Milky Way often have a supermassive black hole (SMBH) surrounded by a dense stellar cluster at their center \citep[e.g.][]{FerrareseFord05,KormendyHo13,Neumayer+20}. The proximity of the Milky Way's Galactic center presents a unique opportunity to learn about the stellar populations in extremely dense, dynamic environments \citep[e.g.,][]{schodel+03,Ghez+05,Ghez+08,Gillessen+09,Gillessen+17}. Observations of this region have presented a number of astrophysical puzzles, including mysterious, dust-enshrouded stellar objects \citep[e.g.,][]{Ciurlo+20}, a top heavy initial mass function (IMF) \citep[e.g.][]{Bartko+09,Lu+09}, and young-seeming massive stars in the vicinity of the SMBH \citep[e.g.,][]{Ghez+05}. 

Complementing direct observations of the Milky Way's galactic center, tidal disruption events (TDEs) can inform our understanding of the stellar populations in other galactic nuclei. A TDE occurs when a star passes within the tidal limit of a SMBH and becomes ruptured by tidal forces \citep{Hills1975,Rees1988,Alexander99,MagorrianTremaine99,WangMerritt04,macleod_tidal_2012}. During this process, the matter being accreted by the SMBH releases large amounts of energy, producing a bright electromagnetic transient \citep[e.g.,][]{Guillochon+13, Gezari21}. Spectra from TDEs can constrain the composition and mass of the disrupted star, providing a probe of the stellar population at large \citep[e.g.,][]{kochanek_abundance_2016, kochanek_tidal_2016, Yang17,Mockler+22,Miller+23}. With the advent of transient surveys like the Zwicky Transient Survey and the Legacy Survey of Space and Time (LSST) \citep{ZwickySkySurvey+19,van_velzen_optical-ultraviolet_2020,LSSTScienceBook}, TDEs will give us an unprecedented glimpse into the stellar populations in other galactic nuclei.

Direct stellar collisions can shape the stellar populations within the sphere of influence of the SMBH. For example, collisions between stars can alter their orbits and properties, such as their mass and remaining main-sequence lifetime \citep[e.g.,][]{Rauch99,FreitagBenz02,Rose+23}. Previous work has connected them to observations of the Milky Way's galactic center, as well as TDEs of stripped stars \citep[e.g.,][]{Mastrobuono-Battisti+21,Gibson+24,Rose+23,RoseMockler+25}. Furthermore, very high-speed collisions ($\gtrsim$$500$ km/s) can destroy stars, with the highest-speed collisions ($\gtrsim 10000$ km/s) potentially producing supernovae-like electromagnetic transients \citep{Balberg+13,BalbergYassur23,Brutman+24,Ryu+24a,Ryu+24b}. Stars can also become ejected from the cluster after receiving a velocity kick from a collision \citep{RoseMockler+25}. This wide range of potential observables highlights the need to understand how collisions shape the mass demographics of the stars in galactic nuclei.

The result of an individual collision depends on a variety of parameters, such as the mass of each of the colliding stars, their relative speed, impact parameter, and age, and therefore their structure \citep[e.g.,][]{Lai+93}. The outcome of a stellar collision can be fully understood through hydrodynamical simulations. However, over $10$~Gyr of evolution of a dense stellar cluster, thousands to tens of thousands of collisions can occur \citep[e.g.,][]{Sidhu+26} and performing hydrodynamic simulations for each collision is highly impractical.
Previously, simulations of dense stellar clusters have approximated the outcome of a collision using fitting formulae developed by hydrodynamic studies \citep[e.g.,][]{Lai+93, Rauch99, Rose+23}, but vastness of the parameter space can limit the accuracy of fitting formulae  \citep[e.g.,][]{Freitag+02}. However, recent advances in modeling collision outcomes using machine learning (ML) offer a novel way to bridge the detailed physics of direct collisions with their population-level effects in the cluster \citep{AmaroSeoane25,Rose+25,Prieto+26}. ML tools enable population studies of clusters to sample the entire multidimensional parameter space of collisions beyond the limits of previous prescriptions.

Most significantly, \citet{Prieto+26} utilized a multi-layered neural network to predict collision outcomes based on a grid of $27,720$ hydrodynamic simulations. We present the first use of their ML model, \verb|collAIder|, in a simulation of collisional dynamics in the Galactic center. Specifically, this paper builds upon the code developed by \citet{Rose+22,Rose+23} and \citet{RoseMockler+25} by incorporating \verb|collAIder| to predict stellar collision outcomes in the Galactic center. 
Taking a Monte Carlo approach, we examine the impact of stellar collisions on the stellar population. This paper is organized as follows: 

In Section \ref{sec:models}, we discuss the methods used in this study. Sections~\ref{subsec:cluster_properties} and~\ref{subsec:SD} focus on our treatment of the stellar dynamics and Section \ref{subsec:ML} describes the ML methods used to create and train \verb|collAIder|. In Section~\ref{subsec:initial_conditions}, we outline the initial conditions used in this study. In Section~\ref{sec:resultsUM}, we examine the outcomes of two simulations that assume a uniform mass distribution and compare our results to previous studies. In Section~\ref{sec:resultsMS}, we discuss the results of a simulation with a realistic initial mass function. In Section~\ref{sec:ejected_gas}, we evaluate the gas mass injected into the Galactic center as a result of stellar collisions. Finally, we summarize our findings in Section~\ref{sec:conclusion}.

\section{Methodology} \label{sec:models}

We take a statistical approach to model the evolution of stars within the Galactic center. We utilize the code first developed by \citet{Rose+22,Rose+23} and \citet{RoseMockler+25}, which includes the effects of two-body relaxation and physical collisions. We follow a sample of $4000$ stars embedded in a fixed cluster with properties described in Section~\ref{subsec:cluster_properties} below. In order to model the effects of stellar collisions on the stars, we use \verb|collAIder|, discussed in greater detail in Section~\ref{subsec:ML}. This addition represents the main development in the code compared to previous work, which relied on fitting formulae to SPH simulations from the literature to determine collision outcomes.

\subsection{Galactic Center Properties} \label{subsec:cluster_properties}
In this study, we base our initial conditions and simulation set-up on the Milky Way's Galactic center. We model the mass density of the galactic center as a function of distance from the SMBH using the power law:
\begin{eqnarray} \label{eq:density}
    \rho(r_\bullet) = \rho_0 \left( \frac{r_\bullet}{r_0}\right)^{-\alpha} \ , 
\end{eqnarray}
where $\alpha$ is the slope, taken to be either 1.25 or 1.75 in this work, $r_\bullet$ is the distance from the SMBH, and $\rho_0$ is a normalization factor based on observations. Specifically, \mbox{$\rho_0 = 1.35 \times 10^6 \, M_\odot/{\rm pc}^3$} at $r_0 = 0.25 \, {\rm pc}$ \citep{Genzel+10}. 

The SMBH dominates the gravitational potential within the sphere of influence. The velocity dispersion therefore decreases with distance from the SMBH:
\begin{eqnarray}\label{eq:sigma}
    \sigma(r_\bullet) = \sqrt{ \frac{GM_{\bullet}}{r_\bullet(1+\alpha)}},
\end{eqnarray}
where $M_{\bullet}$ is the mass of the SMBH \citep{Alexander99,AlexanderPfuhl14}. In our simulation, we assume $M_\bullet$ to be \mbox{$4 \times 10^6$~M$_\odot$}, similar to the Milky Way's SMBH \citep[e.g.,][]{Ghez+03}.

\subsection{Stellar Dynamics} \label{subsec:SD}

We utilize the code developed in \citet{Rose+23} and \citet{Rose+25}, which includes two-body relaxation and direct collisions. The relaxation timescale is defined as the time it takes for gravitational encounters with other objects in the cluster to change the velocity of a given star by an order of itself. It is given by
\begin{eqnarray} \label{eq:t_rlx}
t_{\rm rlx} = 0.34 \frac{\sigma^3}{G^2 \rho \langle M_\ast \rangle \ln \Lambda_{\rm rlx}} \ ,
\end{eqnarray}
where $\langle M_\ast \rangle$ is the average stellar mass and $\ln \Lambda_{\rm rlx}$ is the Coulomb logarithm \citep[e.g.,][]{BinneyTremaine,Merritt2013}. Over the course of an orbit, we apply a randomly drawn velocity kick to each star such that $\Delta v_\mathrm{kick}/v_\mathrm{star} \sim \sqrt{P/t_{\rm rlx}}$, 
where $P$ is the orbital period \citep[for the full equations and implementation, see][]{Naoz+22,Rose+22}.

The collision rate experienced by a star 
is set by its orbit within the cluster. It is given by
\begin{eqnarray} \label{eq:t_coll_main_ecc}
     t_{\rm coll}^{-1} &=& \pi n(a_\bullet) \sigma(a_\bullet) \nonumber \\ &\times& \left(f_1(e_\bullet)r_c^2 + f_2(e_\bullet)r_c \frac{2G(M_{\rm coll}+M_{\rm star})}{\sigma(a_\bullet)^2}\right)\ 
\end{eqnarray}
where $f_1(e_\bullet)$ and $f_2(e_\bullet)$ are equations 20 and 21 from \citet{Rose+20}, $G$ is the gravitational constant, $a_\bullet$ is the star's semimajor axis, $n(a_{\bullet})$ is the number density, $r_c$ is the sum of the radii of the colliding stars, $M_{\rm star}$ represents the mass of the sample star, and M$_{\rm coll}$ is the mass of the background star from the cluster with which it collides. We take a statistical approach to collisions: we randomly draw a number and compare it to the probability that a collision will occur over a single orbit, given by $P/t_{coll}$. If a collision occurs, we use \verb|collAIder| to predict its outcome, as described below in Section~\ref{subsec:ML}.

We treat the kinematics of stellar collisions following \citet{RoseMockler+25}. For collisions which result in a merger, we determine the final orbit of the merger product using momentum conservation. On the other hand, some collisions have either too large of an impact parameter or too high of an impact speed for the colliding stars to become bound to one another. These collisions deflect the stars by angle $\Delta \theta$ in their center of mass frame \citep{Rose+26}. Our treatment of these collisions is qualitatively similar to the prescription outlined in \citet{RoseMockler+25}, except we use the more recent fitting formulae to SPH simulations from \citet{Rose+26}, where the deflection angle from the collision is given by
\begin{align}
    \Delta \theta = 2\arctan\left(\frac{b_{90}}{b}\right)\bigg(1 &+ Ae^{-\left(\frac{r_{\rm{p}}}{R_1+R_2}\right)^2} \notag \\
    &- B\frac{v_{\infty}}{v_{\rm{esc}}}e^{-a\left(\frac{r_{\rm{p}}}{R_1+R_2}\right)^2}\bigg)
\end{align}
with \(A = 0.16\), \(B = 0.35\), and \(a = 2.5\), $v_{\infty}$ is the relative velocity of the two stars at infinity, $v_{\rm esc}$ is the escape velocity of the star, \(b_{90}\) is the impact parameter required for a 90\(\degree\) deflection and \(b\) is the impact parameter of the collision, related to the periapsis separation $r_p$ of the collision by
\begin{equation}
    b^2 = r_{\rm{p}}^2 + \frac{2G(M_1 + M_2)r_{\rm{p}}}{v_{\infty}^2}
\end{equation}
We note that \citet{Rose+26} also provide a fitting formulae for the velocity kick magnitudes due to a collision. We reserve the inclusion of this fitting formulae for future work as they were developed for specific mass ratios of stars and, as shown in \citet{RoseMockler+25}, changing how dissipative stellar collisions are does not affect the final mass demographics of the stars. 

We estimate the lifetime of our sample of main-sequence stars as
\begin{equation}\label{eq:lifetime}
\tau = 10\times\left(\frac{M_\odot}{M_{*}}\right)^{2.5} \text{ Gyr.}
\end{equation}
This scaling relation corresponds to solar metallicity, such that the lifetime of a $1$~M$_\odot$ star is roughly $10$~Gyr, the duration of our simulation. Following a collision, we update the lifetime and radius of the star using its new mass \citep[for details, see][]{Rose+23}. If at any point in the simulation, the mass of a star drops below $M_{\rm min} = 0.08$ M$_{\odot}$, the lower limit of a Kroupa IMF \citep[][]{Kroupa01}, we assume the star has been destroyed. We designate any collision that causes the star's mass to fall below $0.08$~M$_\odot$ as a destructive collision (DC).

Because \verb|collAIder| was trained on lower metallicity star, we scale the age of a star from our simulation before feeding it to \verb|collAIder|. For a star with main-sequence lifetime $\tau$ given by Eq.~(\ref{eq:lifetime}), the approximate age of the star at $0.01$ Z$_\odot$ is 
\begin{equation} \label{eq:scaled_lifetime}
    t' = \frac{t}{\tau}\times \tau',
\end{equation}
where t is the current age of the star and $\tau'$ is lifetime scaling relation for $0.01$ Z$_\odot$, given by
\begin{equation} \label{eq:low_lifetime}
\tau' = 0.527\times10^{3.98\left(M_*/M_\odot\right)^{-0.371}}{ \rm Myr}.
\end{equation}
We obtained this scaling relation by fitting the terminal age main-sequence (TAMS) ages from the specific \verb|MESA| models used in \verb|collAIder|'s training data \citep{Prieto+26}. We also estimate the radius of a star using
\begin{equation} \label{eq:radius}
    R_{*} = R_\odot \times \left(\frac{M_{*}}{M_\odot}\right)^{0.57}.
\end{equation}

At each timestep in our simulation, we check if a star's orbit carries it close enough to the SMBH to become ruptured by tidal forces. This critical distance is given by
\begin{eqnarray} \label{eq:roche}
    R_\mathrm{tidal} = R_{\rm star}\left(2\frac{M_{\bullet}}{M_{\rm star}}\right)^{1/3}.
\end{eqnarray}
where $R_{\rm star}$ and $M_{\rm star}$ are the radius and mass of the sample star, respectively \citep[e.g.,][]{Chandrasekhar+63, macleod_tidal_2012, Guillochon+13}.  If the periapsis of the star's orbit is less than this critical radius, we determine that a TDE has occurred and destroy that star. Otherwise, we allow a star to continue in the simulation until it has reached the end of its main-sequence lifetime, as described above, or our desired runtime of $10$~Gyr, whichever comes first. 

\subsection{Machine Learning Model to Predict Collision Outcomes} \label{subsec:ML}

We use \texttt{collAIder}, a publicly available tool, to predict stellar collision outcomes and remnant properties \citep{Prieto+26}. \texttt{collAIder} is trained on a grid of $27,720$ SPH simulations of collisions involving main-sequence stars. The simulations were performed using \texttt{StarSmasher} \citep{Rasio91, Gaburov+10}, which used realistic \texttt{MESA} stellar profiles \citep[v24.08.1;][]{Paxton+11, Paxton+13, Paxton+15, Paxton+18, Paxton+19, Jermyn+23} based on the parameters described in \verb|POSYDON| \citep[v1;][]{Fragos+23}, neglecting winds.

The base grid of SPH collisions covers stellar masses from $0.2$ \(M_{\odot}\) to $64$ \(M_{\odot}\) at ages ranging from $0.001$ Gyr to $13.7$ Gyr, all with a metalicity of $0.01$ $Z_{\odot}$. Stars collide at relative velocities at infinity ranging from $10$ to $16000$ km/s and at pericenter distances sampled from directly head-on to a grazing encounter at the sum of both stellar radii. A detailed overview of the SPH grid, as well as a comparison of different ML techniques can be found in \cite{Prieto+26}. 

\texttt{collAIder} uses two independently trained neural networks (NNs), with one performing classification and the other regression. The classification task predicts the number of stellar remnants after the collision. There are three possible outcomes of a stellar collision, corresponding zero, one, or two stars surviving. There is also nuance in the one remnant case: either the stars will experience a merger, or one star will be completely disrupted while the other survives, potentially losing a large fraction of its mass in the process. Therefore, \texttt{collAIder} defines four collision outcomes with the labels $0$, $1$, $2$, representing destruction of both stars, a merger, and two  surviving stars, respectively, and label $3$ representing a collision that results in the destruction of one star while the other survives. \texttt{collAIder} achieves a test balanced accuracy of $98.2\%$. 

The regression task predicts the final mass of the surviving stars. To ensure mass conservation is obeyed, the NN predicts fractional masses as shown below:
\begin{equation} \label{eq:norb}
\biggl\{\frac{M_{1,f}}{M_{\mathrm{tot},i}}, \frac{M_{2,f}}{M_{\mathrm{tot},i}}, \frac{M_{u,f}}{M_{\mathrm{tot},i}}\biggr\}
\end{equation}
where \(M_{1,f}\), \(M_{2,f}\) are the final masses of stars 1 and 2, respectively, \(M_{u, f}\) is the unbound mass, and \(M_{\mathrm{tot}, i}\) is the total initial mass. A softmax function is applied to the network outputs, which enforces mass conservation by ensuring that the three fractions sum to one. When a merger occurs, the final mass is always assigned to \(M_{1,f}\). \texttt{collAIder} predicts the final stellar masses with test median relative errors below $0.15\%$. Throughout this paper, we refer to all stars that lose more than $10\%$ of their initial mass as ``stripped stars.''

\subsection{Initial Conditions} \label{subsec:initial_conditions}
We run two simulations in which all of the sample stars are initially $1$ M$_{\odot}$. Similarly, all stars in the background cluster, where the colliding stars are drawn from, are fixed at 1 M$ _{\odot}$. We consider two possibilities for the stellar density profile: with $\alpha = 1.75$, consistent with a Bachall-Wolf distribution \citep[][]{BahcallWolf76}, and $\alpha = 1.25$, based on Fokker-Plank models that include a population of stellar-mass BHs \citep{DuncanShapiro83,Murphy+91,David+87a,David+87b,FreitagBenz02,AharonPerets16,LinialSari22, Rose+23, RoseMacLeod24}. These choices encapsulate the range of density profiles suggested by theoretical and computational studies as well as observations of the Galactic center \citep{Schodel+18,Schodel+20,Gallego+18}.

We also run a simulation with an $\alpha = 1.75$ density profile and a top-heavy IMF ranging from $M_{\rm{min}} = 0.5$ M$_{\odot}$ to $M_{\rm{max}} = 30$ M$_{\odot}$, based on observations of the Galactic center \citep{Lu+13}. 
If a collision occurs, we assume the sample star collides with a typical star from the background cluster. Its mass is the average stellar mass within the cluster at the time of collision, given an IMF. Observations of the Galactic center suggest that the bulk of the stars ($\sim 90\%$) formed in a single star formation episode \citep[e.g.,][]{Chen+23}. We therefore assume that all of the stars in our simulation formed at $t=0$ and have the same age.

We initialize 4000 stars with orbital eccentricity drawn from a thermal distribution, expected for a dynamically relaxed system \citep[][]{RauchTremaine96,HopmanAlexander06,Merritt10,Merritt2013}. We draw the semi-major axes of the orbits so that the stars lie on a cusp given by the stellar density profile (Eq.~(\ref{eq:density})) between \mbox{$0.001$ pc} and 1 pc from the SMBH. 
We assume that the cluster is isotropic and spherically symmetric.  

\begin{deluxetable*}{lcccc}
\tablecaption{Summary of key outcomes for the three cluster models.\label{tab:model_summary}}
\tablewidth{0pt}
\tablehead{
  \colhead{Quantity} &
  \colhead{UM $\alpha=1.75$} &
  \colhead{UM $\alpha=1.25$} &
  \colhead{IMF $\alpha=1.75$} &
  \colhead{Section(s)}
}
\startdata
\sidehead{\textit{General}}
Total sampled stars (within $1$ pc of SMBH) & 4000 & 4000 & 4000 & \ref{sec:resultsUM}, \ref{sec:resultsMS} \\
Sampled stars within $0.1$ pc of SMBH & 225 & 73 & 192 & \ref{sec:resultsUM}, \ref{sec:resultsMS} \\
Sampled stars within $0.01$ pc of SMBH & 12 & 0 & 7 & \ref{sec:resultsUM}, \ref{sec:resultsMS} \\
Total collisions to sample stars                        & 14{,}806      & 6{,}415       & 3{,}500       & \ref{sec:resultsUM}, \ref{sec:resultsMS} \\
Stars with $\geq 1$ collision [\%]                      & 39.00          & 37.67         & 19.13          & \ref{sec:resultsUM}, \ref{sec:resultsMS} \\
Stars with $>10\%$ mass loss [\%]                       & 18.81          & 11.18          & 7.40          & \ref{sec:resultsUM}, \ref{sec:resultsMS}, \ref{sec:conclusion} \\
Stars with $>50\%$ mass loss [\%]                       & 16.96          & 8.23           & 6.29          & \ref{sec:resultsUM}, \ref{sec:resultsMS}, \ref{sec:conclusion} \\
Stars ejected from cluster [\%]                            & $2.28$     & 1.89          & $1.23$     & \ref{sec:resultsUM}, \ref{sec:resultsMS} \\
Stars experiencing a TDE [\%]                           & $0.53$     & $0.66$    & $0.21$    & \ref{sec:resultsUM}, \ref{sec:resultsMS} \\
\sidehead{\textit{Mergers}}
Collisions resulting in a merger [\%]                   & 1.74          & 2.67          & 3.08          & \ref{sec:resultsUM}, \ref{sec:resultsMS} \\
Stars that experienced a merger [\%]                           & 6.47          & 4.30          & 2.85          & \ref{sec:resultsUM}, \ref{sec:resultsMS} \\
Stars with net mass gain [\%]                           & 4.82          & 3.90          & 6.37          & \ref{sec:resultsUM}, \ref{sec:resultsMS} \\
\sidehead{\textit{Destructive Collisions (DCs)}}
Collisions resulting in a DC [\%]                       & 4.33          & 4.01           & 6.60           & \ref{sec:resultsUM}, \ref{sec:resultsMS} \\
Stars destroyed in a DC [\%]                            & $16.10$    & $6.54$    & $6.10$    & \ref{sec:resultsUM}, \ref{sec:resultsMS} \\
Avg.\ collisions before DC                              & 18            & 10            & 9             & \ref{subsec:DC_Properties}, \ref{sec:resultsMS} \\
Avg.\ DC relative velocity [km s$^{-1}$]              & 5{,}733       & 6{,}580       & 5{,}612       & \ref{subsec:DC_Properties}, \ref{sec:resultsMS} \\
Avg.\ DC normalized periapsis separation                & 0.30          & 0.30          & 0.22         & \ref{subsec:DC_Properties}, \ref{sec:resultsMS} \\
\sidehead{\textit{Ejected Gas}}
Total ejected mass [\% of initial stellar mass]         & 18.89         & 9.13          & 2.11           & \ref{sec:ejected_gas} \\
Total ejected mass [M$_\odot$ scaled to cluster] & $755,600$ & $365,200$ & $251,252$ & \ref{sec:ejected_gas} \\
Ejected mass retained by SMBH [\%]                     & $90.83$    & \nodata       & $88.35$    & \ref{sec:ejected_gas} \\
Global injection rate [M$_\odot$/yr]                     & $7.51\times10^{-5}$     & \nodata       & $3.17 \times 10^{-5}$    & \ref{sec:ejected_gas} \\
\enddata
\end{deluxetable*}

\section{Results of the Uniform Mass Models} \label{sec:resultsUM}

\subsection{General Outcomes} \label{subsec:gen_outcomes_UM}

A summary of key outcomes for all three simulations is provided in Table~\ref{tab:model_summary}. We begin by examining the results of the two uniform mass models with all $1$~M$_\odot$ stars and density profile slopes of $\alpha =1.75$ and $\alpha = 1.25$, respectively. In both simulations, about $40\%$ of stars experienced at least one collision. However, the simulation with the steeper cusp ($\alpha = 1.75$) had a higher number of collisions, $14806$ compared to $6415$ in the $\alpha = 1.25$ model. This difference is a consequence of the spatial distribution of the stars. The simulation with the steeper cusp had three times the number of stars in the inner $0.1$~pc region compared to the $\alpha = 1.25$ case ($225$ versus $73$ out of $4000$ sample stars).
However, the qualitative outcomes of collisions were similar in both cases. Out of the total number of collisions within a simulation, \verb|collAIder| predicted that $2$ to $3\%$ resulted in a merger, while $4\%$ were a DC. The remaining collisions, the overwhelming majority, were predicted to be hit-and-run collisions (i.e. collisions that result in two surviving stars).

\begin{figure*}[t!]
\centering
\includegraphics[width=0.49\textwidth]{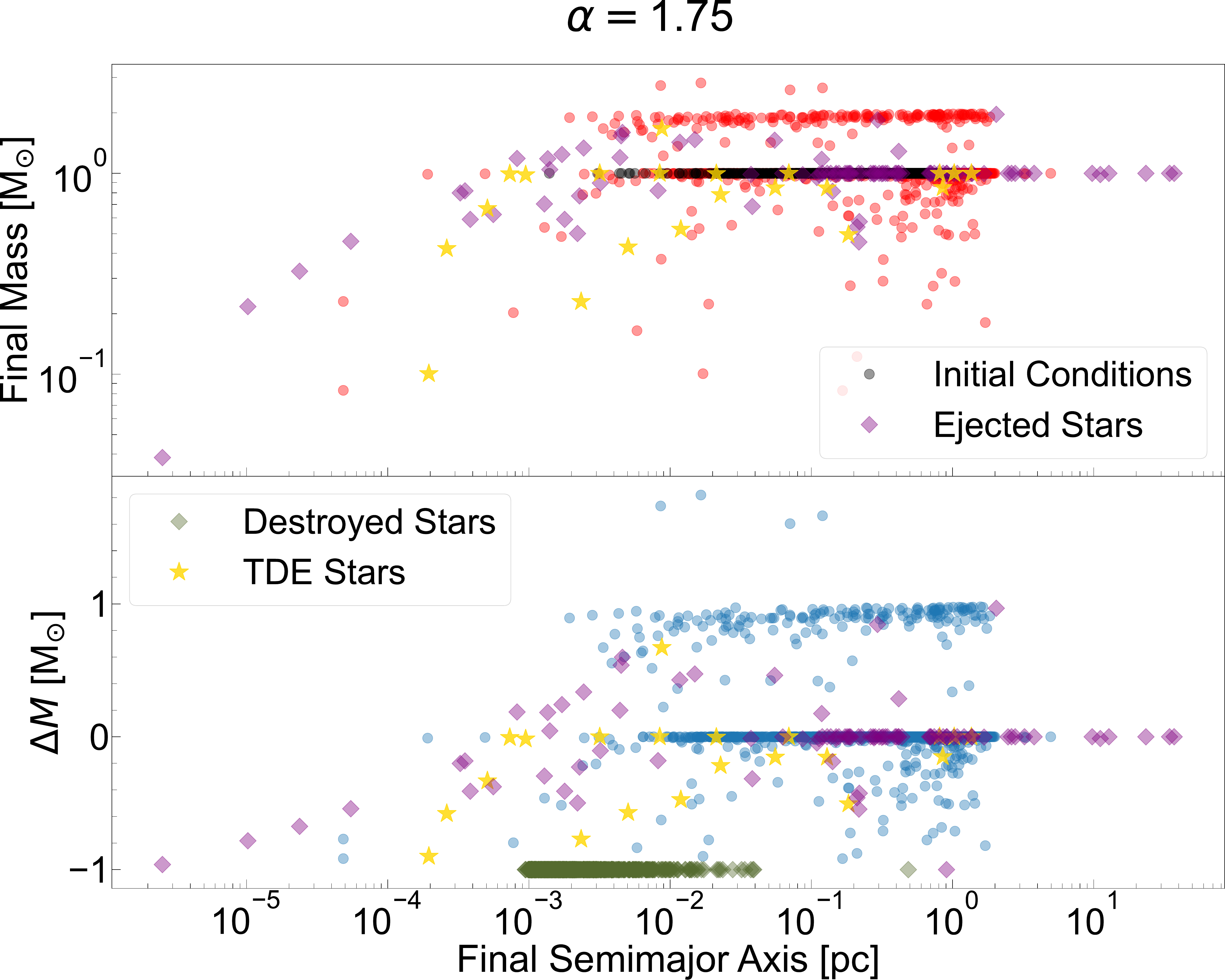}\hfill
\includegraphics[width=0.49\textwidth]{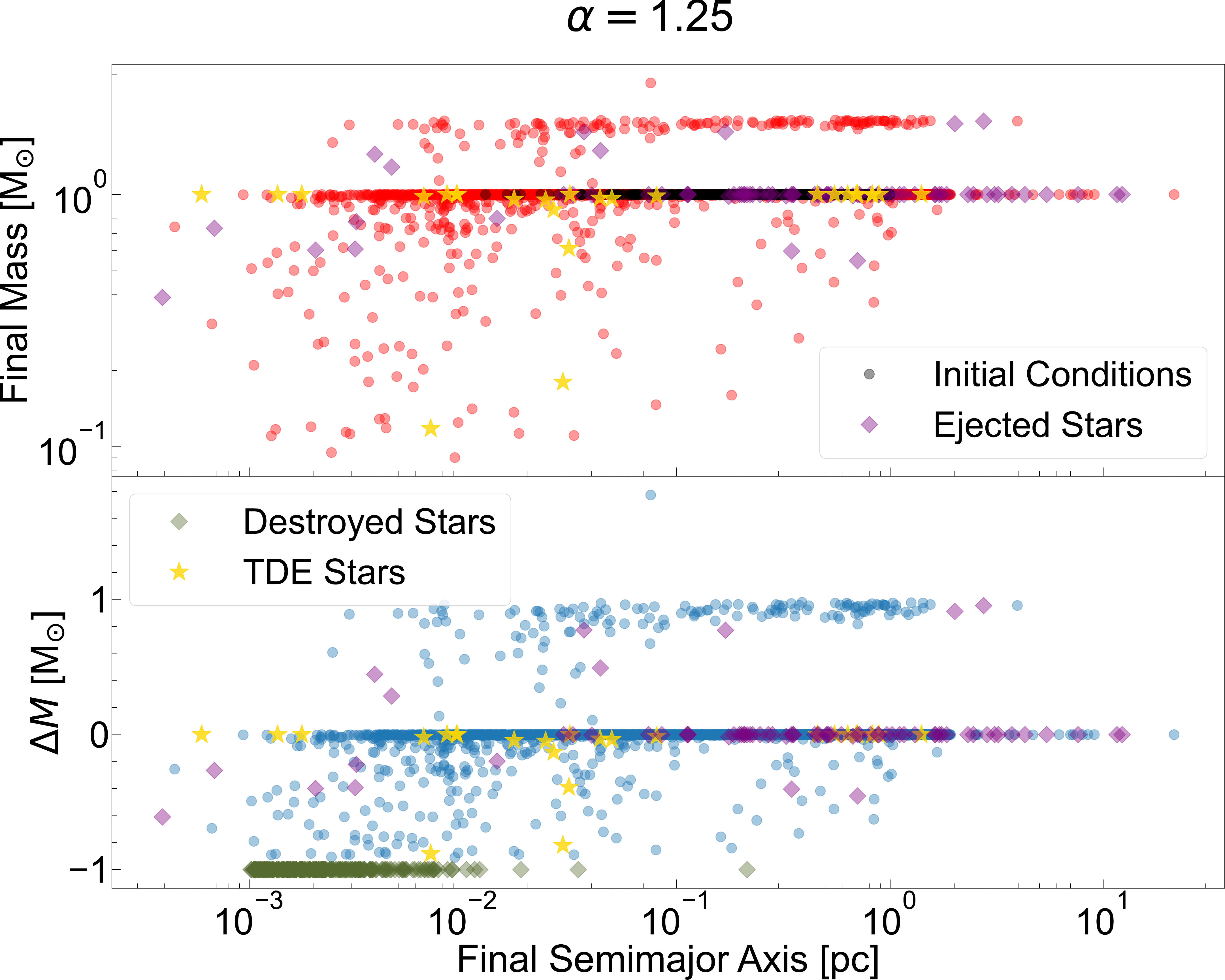}
\caption{This figure shows the final mass of a star versus its final semimajor axis from the SMBH for the $\alpha=1.75$ (left) and $\alpha=1.25$ (right) uniform mass models with 1M$_\odot$ stars. The final values are represented in red (top) and blue (bottom). Initial conditions of the stars are presented in black in the top panels. Around $40\%$ of stars experience a collision in both models, with most being stripped of some mass. Any stars that lost all of their mass (bottom panels with $\Delta M = -1$ M$_{\odot}$) are DC stars, represented by green diamonds. Stars ejected from the cluster are pictured in the top panels by purple diamonds and stars that experience a TDE are shown as gold stars in both panels. We note that for the disrupted stars, the position and mass are taken from the orbital parameters and present mass of the star and when it is removed from the simulation. For ejected stars, we present the absolute value of their final semimajor axis.}
\label{fig:mass_change}
\end{figure*}

We examine the final masses of the sample stars in Figure \ref{fig:mass_change}. 
The majority of stars do not collide and therefore do not change mass during the simulation; we assume that there is no mass loss from stellar winds, which are thought to be negligible for main-sequence stars of these masses \citep[][]{Johnstone+15a, Johnstone+15b}. Of those stars that experienced a collision in the $\alpha=1.75$ simulation, the majority lost mass: $19\%$ of stars lost at least $10\%$ of their initial mass and $17\%$ of stars lost at least $50\%$, mostly dominated by stars that experienced a DC. These numbers decrease by roughly a factor of two for the $\alpha = 1.25$ case.  

There are fewer DCs in the $\alpha = 1.25$ case compared to $\alpha = 1.75$ for the same reasons there are fewer collisions overall. The shallower cusp leads to a smaller number of stars within the densest, most collision-prone region of the cluster, near the SMBH, where the velocity dispersion is also higher. For a cusp with $\alpha=1.25$, the collision timescale is longer within the inner $0.1$~pc region. DCs are primarily a result of high-speed, nearly head-on collisions near the SMBH (see Section~\ref{subsec:DC_Properties} for more details). Furthermore, they are often preceded by multiple hit-and-run collisions that either chip away at the mass of the star and make it easier to disrupt or are too grazing to affect the star at all.

We also note that a handful of stars diffuse or are scattered to larger orbits about the SMBH, outside of the sphere of influence (within $1$ pc of the SMBH). Furthermore, grazing hit-and-run collisions can place stars on unbound orbits about the SMBH \citep[see][]{RoseMockler+25}. Some of these stars might escape from the cluster, while others will remain bound to the star cluster even if they are gravitationally unbound from the SMBH. We record these stars as ejected from the inner pc, but reserve a more detailed examination of their fate for future work. In both simulations, $2\%$ of the stars became unbound from the SMBH, slightly higher than previous work \citep[$0.7\%$ in][]{RoseMockler+25} owing to the updated kinematics fitting formulae. This value represents an upper limit to the number of ejected stars and may change by up to 50\% due to dissipation during the collisions \citep{RoseMockler+25}.

\subsection{Mass Demographics}

\begin{figure*}[htb!]
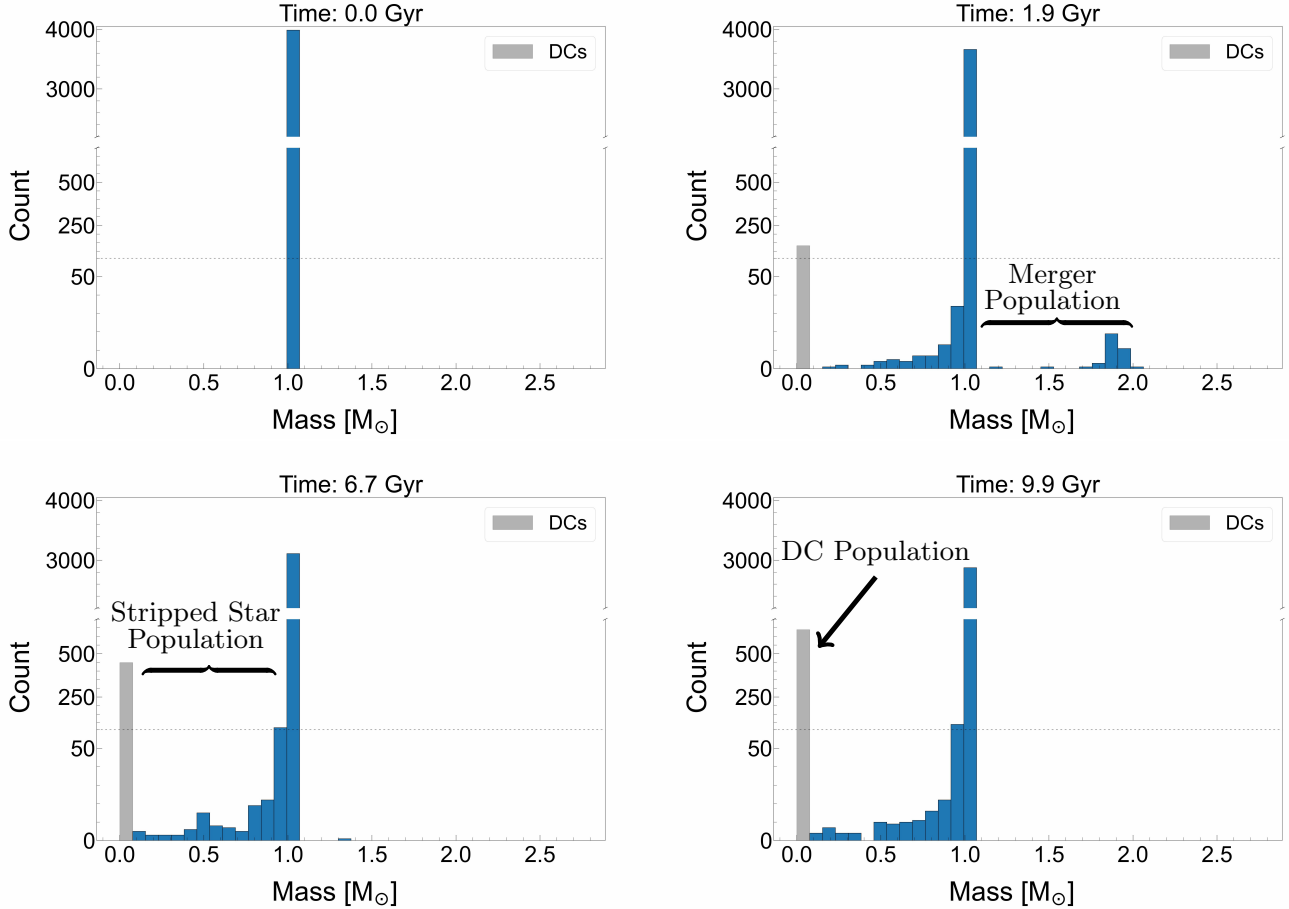

\begin{center}

\resizebox{\textwidth}{!}{%
\begin{tikzpicture}[ultra thick]
    \node[anchor=south west, inner sep=0] (img) at (0,0) {
        \plottwo{PaperPlots/mass_evolution_UM_frame_0.pdf}{PaperPlots/mass_evolution_UM_frame_1.pdf}
      };
      \begin{scope}[x=1cm,y=1cm]
    \draw [decorate,
    decoration = {calligraphic brace}] (11.3,1.3) --  (13.0,1.3);
    \node[above] at (12.1,1.6) {Merger};
    \node[above] at (12.1,1.3) {Population};
    
  \end{scope}
\end{tikzpicture}%
}


\resizebox{\textwidth}{!}{%
\begin{tikzpicture}[ultra thick]
    \node[anchor=south west, inner sep=0] (img) at (0,0) {
        \plottwo{PaperPlots/mass_evolution_UM_frame_2.pdf}{PaperPlots/mass_evolution_UM_frame_3.pdf}
      };
      \begin{scope}[x=1cm,y=1cm]
      
    \draw[->]        (10.1,3.8)   -- (9.45,3);
    
    \node[above] at (10.1,3.8) {DC Population};
    
    \draw [decorate,
    decoration = {calligraphic brace}] (1.8,2.7) --  (3.3,2.7);
    \node[above] at (2.4,3.1) {Stripped Star};
    \node[above] at (2.4,2.8) {Population};
  \end{scope}
  
\end{tikzpicture}%
}

\end{center}

\caption{Mass distribution for the $\alpha=1.75$ model over time. As seen, a large majority of stars do not experience a collision and thus remain 1 M$_{\odot}$ over the course of the entire simulation. As the simulation progresses, we see a number of stars experiencing collisions that cause them to progressively lose mass, denoted as the stripped star population. The gray population represents stars that experienced DCs, i.e. stars with a final mass of 0 M$_{\odot}$. This plot is animated in the digital copy of the paper.}
\label{fig:mass_evolution}
\end{figure*}

We examine the mass distribution of the stars, as shaped by collisions, at snapshots in time in Figure \ref{fig:mass_evolution}. 
For our initially uniform population of stars, most of the sample stars do not collide and therefore remain $1$~M$_\odot$ throughout the simulation, producing a peak at this value in Figure \ref{fig:mass_evolution}. Of the stars that experience at least one collision, a large portion ($\backsim$$70\%$) will go on to experience another collision and sometimes many more. Stars that experience multiple collisions while on the main-sequence can lose mass incrementally over time.

This behavior can be seen in Figure \ref{fig:mass_evolution} to the left of the 1 M$_{\odot}$ peak; a decaying number of stars approach lower masses. Mergers with minimal mass loss produce a peak around $2$~M$_\odot$ in the upper right plot. These merged stars can also lose portions of their mass through later stripping collisions, populating the gap between $1$ and $2$~M$_\odot$.
We find qualitatively similar results for the $\alpha = 1.25$ simulation, shown in Appendix \ref{app:1p25Results}, although it has a larger stripped star population with approximately $80\%$ more stripped stars. As described above, this increase in stripped stars is a result of fewer collisions overall: there are fewer collision sequences that lead to DCs. 

\subsection{Destructive Collisions} \label{subsec:DC_Properties}

\begin{figure}[t!]
\hspace{-1.22em}\includegraphics[width=1\columnwidth]{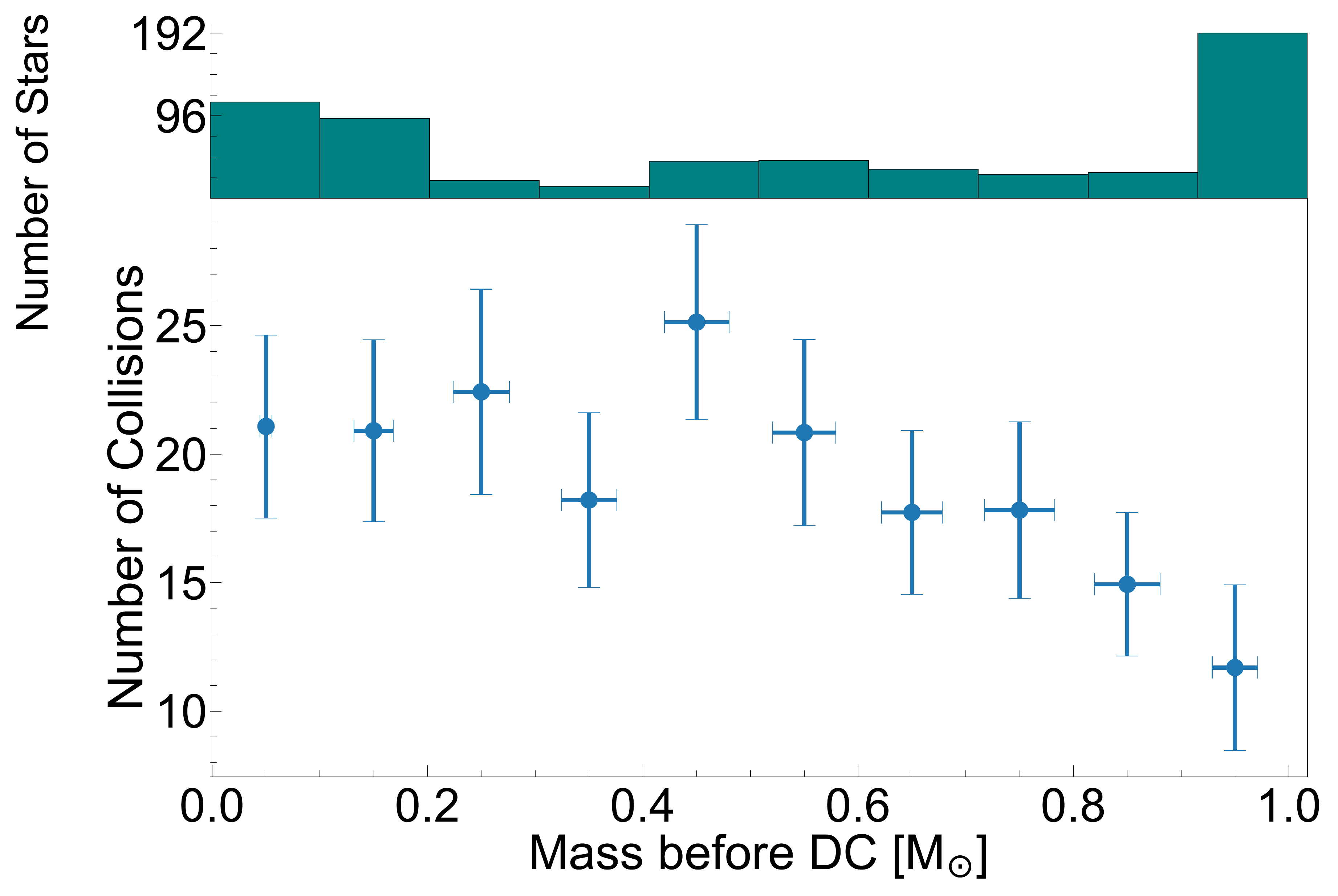}
\includegraphics[width=1\columnwidth]{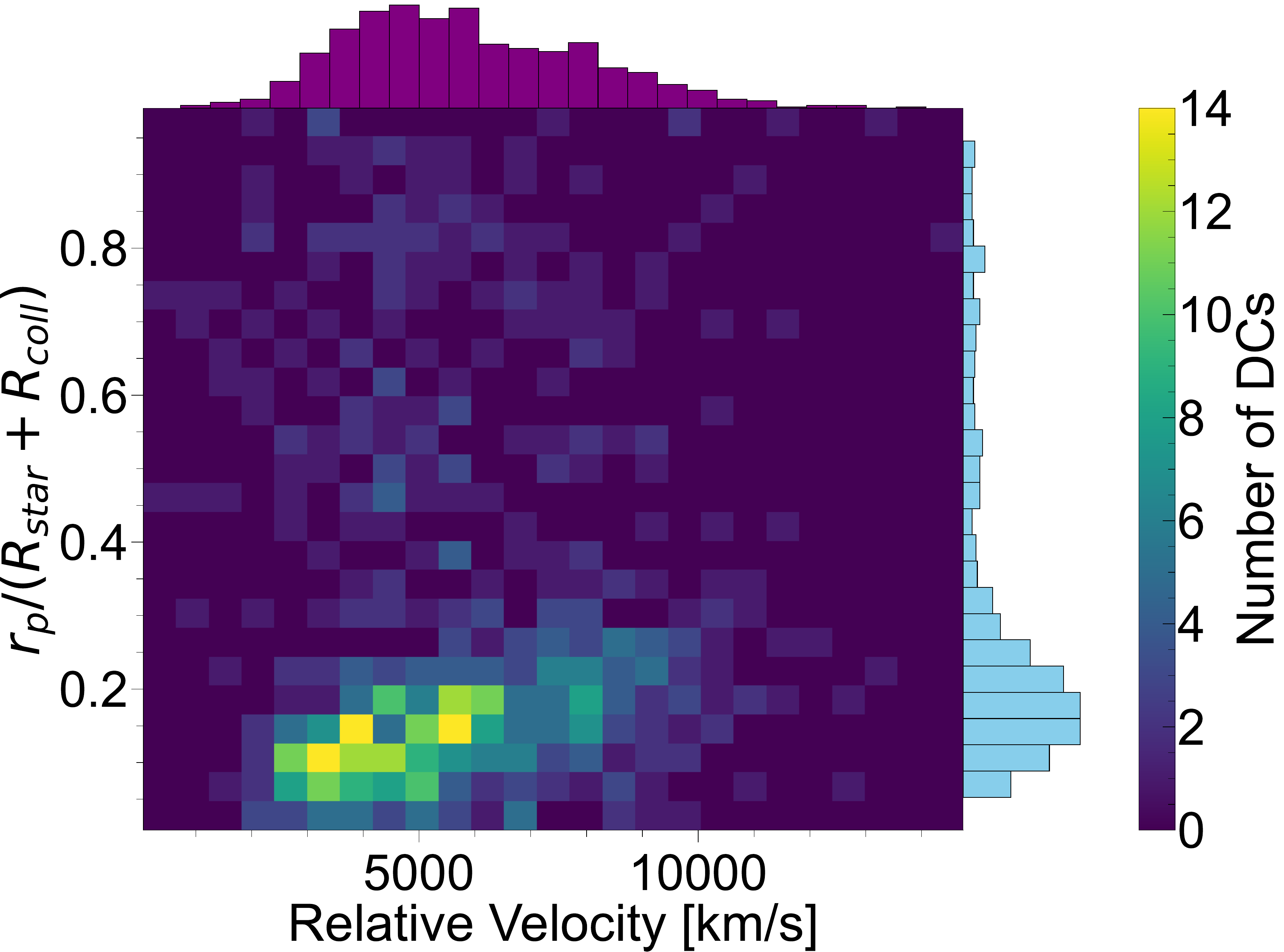}
\caption{\textbf{Mass of stars prior to destruction (Top):} This figure shows the mass of DC stars prior to their final collision for a uniform mass model with an $\alpha=1.75$ stellar density profile. In the top of this plot, we see a count of the number of stars at each mass. In the bottom of this plot, we see the average number of collisions for each mass bin. The error in the number of collisions is determined by $\pm\sqrt{\sigma_{count}}$ and the deviation in the mass is determined as $\pm\sigma_{M}$ in mass. \textbf{DC properties (Bottom):} This figure illustrates the conditions that lead to DCs for the $\alpha=1.75$ model. As seen in the figure, most DCs occur at speeds of at least 2000 km/s, typically nearly head-on. We see that the average conditions for a DC is a very high speed collision of approximately $5700$ km/s with a normalized impact parameter of $0.3$.}
\label{fig:DC_Params}
\end{figure}

In the $\alpha = 1.75$ model, $16\%$ of the stars are destroyed by collisions. In the bottom panels of Figure \ref{fig:mass_change}, which show the net change in mass $\Delta M$ of each star versus its final semimajor axis, the green points at $-1$~M$_\odot$ signify the collection of stars that experienced a DC. The top panel of Figure \ref{fig:DC_Params} shows the mass of each of these stars right before their destruction versus the number of collisions that occurred over their lifetimes. We find that in most cases a star experiences at least 10 collisions prior to the DC, with the average being $18$ collisions. 

The bottom panel of Figure \ref{fig:DC_Params} shows the relative velocity at infinity and normalized periapsis separation of DCs. These collisions tend to be very fast (average relative speed of $5700$ km/s) and nearly head-on (average periapsis separation, normalized to the radii of the two colliding stars, of $0.3$). These collisions occur within $0.01$ pc of the SMBH, where typical velocities exceed thousands of kilometers per second. The distribution of normalized periapses have a tail extending to values $\gtrsim 0.3$. However, these stars have very low masses prior to the DC, already approaching the minimum star mass of $0.08$ M$_{\odot}$ after losing most of their mass through prior stripping collisions. Stars within $5\%$ of $1$~M$_\odot$ that experience a DC had a small average normalized periapsis separation of $0.17$.

For the $\alpha = 1.25$ model, $7\%$ of stars experienced a DC, with a minimum of 5 collisions prior to destruction. On average, these stars experienced 10 collisions prior to destruction. 
The DCs have an average relative velocity of $6600$ km/s and an average normalized periapsis of $0.3$. More energetic collisions are required to disrupt the stars as they had fewer prior collisions to chip away at their mass. 

\subsection{Mergers}

About $5\%$ of the sample stars experienced a merger in both the $\alpha = 1.25$ and $\alpha = 1.75$ models. However, we observe that the majority of stars that gain mass through mergers subsequently lose at least some of that mass during destructive, stripping collisions. In fact, not all stars that merged had a net mass gain by the end of their lifetime, though the majority, $75\%$ to $90\%$, do (see Table~\ref{tab:model_summary}). Scaling our results to the number of stars in the inner pc of the Galactic center, we estimate that $\sim 10^5$ merger products should form over $10$~Gyr. 

Mergers are a less frequent outcome in these simulations due to the high velocity dispersion in the cluster ($>100$~km/s in the inner pc). This reason, coupled with the fact that merger products can still undergo subsequent stripping collisions, mean that stars rarely grow to more than double their masses. The stars in our simulation experienced at most two mergers, with a maximum final stellar mass of $2.8$~M$_\odot$. We note that we only follow a sample of $4000$ stars, while in the Galactic center, the number of stars in the inner pc region is closer to $4 \times 10^6$. It is possible that more massive merger products form through collision sequences, but we are unable to resolve these more rare outcomes with our sample size. However, we estimate that only about $10^3$ stars experience two or more mergers in the inner pc region.

Mergers may be connected with different observables. For example, the magnetic field energies of merger products can be enhanced by 9-12 orders of magnitude, which has important implications for TDEs (see Section~\ref{sec:TDE_implications_UMM}) \citep{Wickramasinghe+14, Schneider+16, Schneider+19, Schneider25, Frost+24, Vynatheya+26}. Furthermore, it has been suggested that luminous red novae (LRNe) can result from stellar mergers \citep{Mason+10, Tylenda+11, Matsumoto&Metzger22}. Across all of our simulations, we have typical merger rates around $10^{-5}$ yr$^{-1}$. However, due to the short observational timescales of LRNe ($\backsim$$10$ years), it is unlikely that we will be able to observe any formed from stellar mergers in our Galactic center. 

While the merger rate is low, the overall collision rate is much higher, at $>10^{-4}$ per year. Peculiar dust- and gas-enshrouded objects have been observed in the Milky Way's Galactic center, known as the G objects \citep{Ciurlo+20}. Stellar collisions represent a possible formation channel for these objects: immediately after the impact, stars may be quite puffy and distended \citep{Rose+23,Gibson+24}. 
If the stars take $\sim 10^4$ years to relax after the collision, we should observe at least $1$ G object from direct collisions in the Galactic center.

\subsection{Tidal Disruption Events} \label{sec:TDE_implications_UMM}

In the $\alpha = 1.75$ model, $\lesssim$$1\%$ of stars experienced a TDE. Of the stars that experienced a TDE, 
about half were stripped in a prior collision(s). The average mass of a star that underwent a TDE was approximately $0.8$ M$_\odot$. The prevalence of stripped stars in TDEs was slightly lower in the $\alpha = 1.25$ simulation, falling to roughly $1$ out of $10$ stars. Stripped stars that eventually experience a TDE may display unique observational signatures that are expected to be hydrogen- and carbon-poor and helium- and nitrogen-rich \citep{Gibson+24}.


Furthermore, one of the disrupted stars from the $\alpha = 1.75$ simulation was a merger product of a collision. As previously noted, merger products may have enhanced magnetic fields \citep{Wickramasinghe+14, Schneider+16, Schneider+19, Schneider25, Frost+24, Vynatheya+26}. 
These enhanced magnetic fields may be involved in fueling observed x-ray emission from TDE sources \citep{Shakura&Sunyaev73, Mummery&Balbus20, Pacuraru+26} and widening the unbound stream of the TDE, increasing the brightness of radio signatures \citep{Yalinewich+19, Lu&Bonnerot20, Pacuraru+26}.
Furthermore, \citet{Bradnick+17} examine mergers from stellar binaries, a different mechanism than we consider here, and suggest that the disruption of merger products by a SMBH may explain the prompt formation of a relativistic jet in some TDEs, though this mechanism is still uncertain \citep[e.g.,][]{Tripto+26}.



\subsection{Comparison to Previous Prescriptions}

We compare our simulations using \verb|collAIder| to previous prescriptions based on fitting formulae and heuristic arguments, with and without two-body relaxation ``turned on'' in the code. We assume the same initial conditions as discussed in Sections~\ref{subsec:initial_conditions} and \ref{subsec:gen_outcomes_UM}. We summarize these results, alongside comparisons to previous prescriptions, in Table~\ref{tab:previous_studies}.


Compared to \verb|collAIder|, the \citet{Rauch99} prescription over-predicts the rate of mergers. The prescription based on \citet{Rauch99} predicts nearly a quarter of all collisions to be mergers and results in more than twice as many stars with a net mass gain. Crucially, \citet{Rauch99} did not include a fitting formula for the capture radius, the maximum impact parameter that would result in a merger for a given speed at infinity. Instead, they note that if the initial speed is roughly the escape speed from the star, a merger can occur. This heuristic treatment was implemented alongside their fitting formulae for mass loss in \citet{Rose+23}, whose prescription we use for comparison to \verb|collAIder|. The \citet{Rauch99} mass loss fitting formulae also result in a slightly smaller fraction of stars that are ultimately destroyed. 

Conversely, \citet{Lai+93}, which includes fitting formulae for both the mass loss and capture radius, under-predicts the rate of mergers and results in fewer stars with net mass gain. Additionally, we see fewer stripped stars when applying these fitting formulae. Compared to hydrodynamic simulations of stellar collisions in \citet{Rose+26}, \citet{Lai+93} underpredicts the mass loss from high-speed collisions, which may contribute to the lower rates of stripped stars. We note that fitting formulae can be difficult to formulate in such a way that the encapsulate the full parameter space, as the collision outcome depends on the relative speed, impact parameter, masses, and stellar structure, and therefore the ages and potentially the metallicity of the stars \citep[e.g.,][]{FreitagBenz02}. For simulations with an initial mass function and stellar evolution, AI/ML tools have a distinct advantage over fitting formulae \citep{Rose+26}. 

In simulations without two-body relaxation, the stellar orbits remain fixed. Consequently, a star's fate is strongly tied to its the distance from the SMBH, a proxy for the velocity dispersion. The percent of stars that lose mass or become destroyed are all $\backsim$$1\%$ across all simulations because $\backsim$$1\%$ of stars reside in the inner $0.01$~pc, where speeds are high enough to make more destructive collisions likely. 
However, relaxation effects must be included to accurately model the Galactic center. Relaxation effects are known to increase the number of collisions \citep[e.g.,][]{Sidhu+26} by allowing stars to diffuse in and out of the direct collision-dominated region of the cluster. Therefore, we also consider the \citet{Rauch99} prescription with relaxation effects included. As shown in Table~\ref{tab:previous_studies}, the fitting formulae accurately predicts the amount of stripped stars. However, the prescription becomes less accurate in predicting the rate of DCs, and continues to over-predict the rate of mergers, similar to the simulation without relaxation.

\begin{deluxetable*}{lccc|ccc}
\tablecaption{Comparison of results to previous prescriptions.\label{tab:previous_studies}}
\tablewidth{0pt}
\tablehead{
  & \multicolumn{3}{c}{\textbf{Without Relaxation}} & \multicolumn{2}{c}{\textbf{With Relaxation}} \\
  \cline{2-4} \cline{5-6}
  \colhead{Quantity} & 
  \colhead{\texttt{collAIder}} & 
  \colhead{\citet{Rauch99}} & 
  \colhead{\citet{Lai+93}} & 
   \colhead{\texttt{collAIder}} &
  \colhead{\citet{Rauch99}}
}
\startdata
Total sampled stars (within $1$ pc of SMBH) & 4000 & 4000 & 4000  & 4000 &  4000    \\
Sampled stars within $0.1$ pc of SMBH & 225 & 216 & 235 & 225 &  206   \\
Sampled stars within $0.01$ pc of SMBH & 12 & 13 & 9  & 12 & 9  \\
Stars with $\geq 1$ collision [\%]                     & 15.80         & 15.78         & 15.00   & 39.00   &    38.24      \\
Stars with $>10\%$ mass loss [\%]                       & 1.23          & 1.35          & 1.15  & 18.81    &    18.86    \\
Stars with $>50\%$ mass loss [\%]                       & 0.83           & 0.68           & 0.55   & 16.96  &     15.71    \\
Stars destroyed in a DC [\%]                           & 0.63    & 0.45   & 0.38 & 16.10  & 13.58    \\
Stars with net mass gain [\%]                          & 4.55          & 10.15         & 3.10   &  4.82 &   10.14       \\
Avg.\ collisions before DC                            & 23            & 20            & 15  & 18    &      19       \\
\enddata
\end{deluxetable*}

\section{Results with an Initial Mass Function} \label{sec:resultsMS}

\subsection{General Outcomes}

Next, we consider an initial mass function for the stars as described in Section~\ref{subsec:initial_conditions}. In this simulation, we assume that the stars lie on a mass density cusp with slope $\alpha = 1.75$. $192$ stars have semimajor axes within $0.1$ pc, and $7$ have semimajor axes within $0.01$ pc. $3500$ collisions occurred over the course of the simulation. Of these collisions, \verb|collAIder| predicted a merger $3\%$ of the time and a DC $7\%$ of the time. Similar to Section~\ref{sec:resultsUM}, most collisions ($\backsim$$90\%$) were hit-and-runs, causing only slight mass loss of both stars.

\begin{figure}[h!]
\includegraphics[width=1\columnwidth]{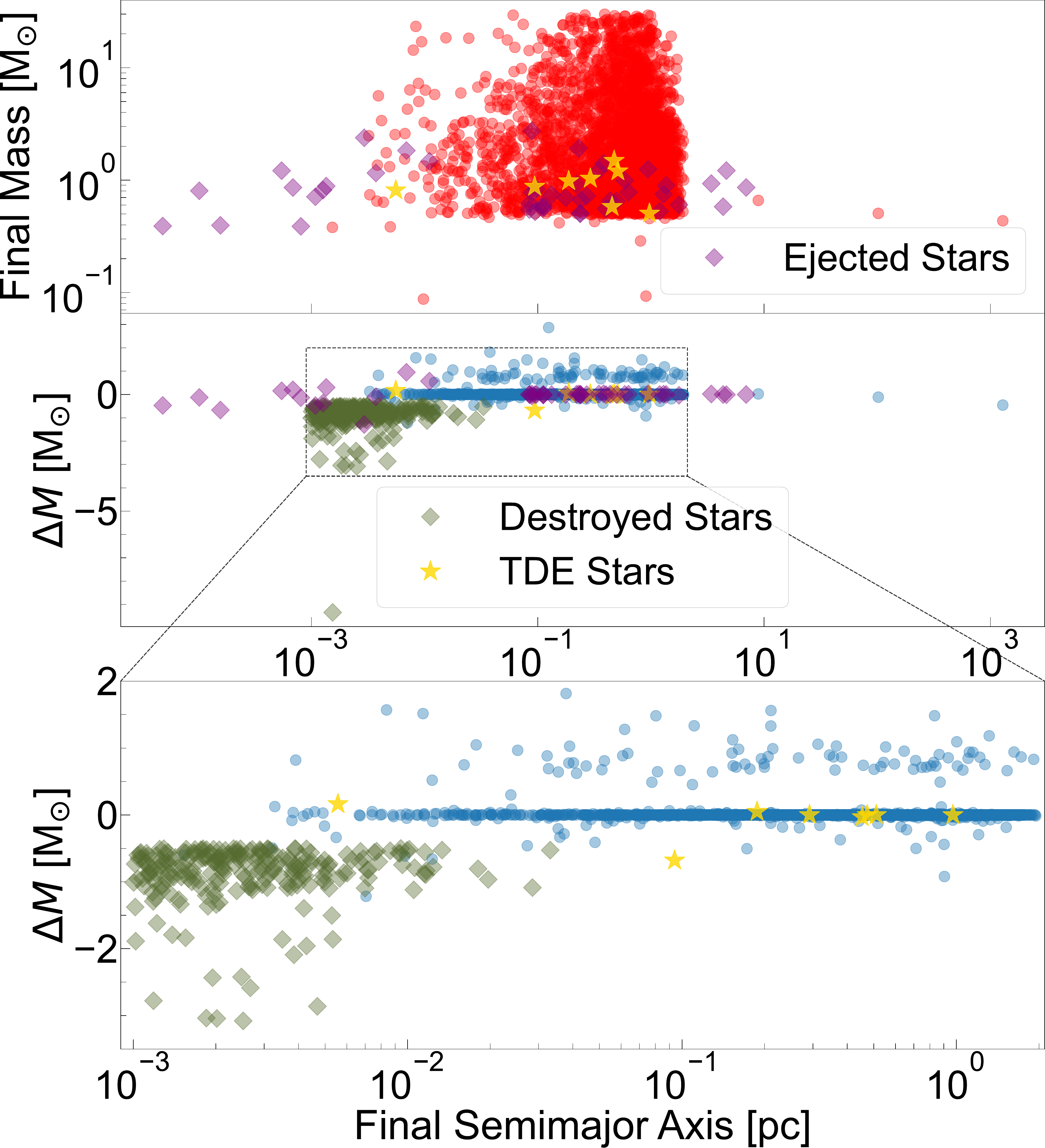}
\caption{This figure shows the final mass of a star versus its final distance from the SMBH for the IMF model. Ejected stars are pictured as purple diamonds, destroyed stars are shown in dark green in the middle and bottom panels, and TDE stars are shown as gold stars in all panels. In the top panel, we see that the most massive stars are never ejected and most stars have very little, if any, mass change. In the middle and bottom panel, we can see the largely decreased stripped star population compared to the uniform mass models. Many stars in this model have incredibly short lifetimes compared to a $1$ M$_\odot$ star which causes them to die before experiencing a collision. A majority of destroyed stars lie incredibly close to the SMBH.}
\label{fig:mass_change_MS}
\end{figure}

In Figure \ref{fig:mass_change_MS}, we examine mass change over the simulation versus the final semimajor axis of each star. Nearly $20\%$ of stars experienced at least one collision, so a large majority of stars had no mass change throughout the simulation. Approximately $7\%$ of stars lost at least $10\%$ of their initial mass and $6\%$ of stars lost at least $50\%$. Of the $4000$ stars sampled, $6\%$ are destroyed in a DC and $1\%$ were placed on an unbound orbit about the SMBH. The lower rates of DCs and collisions in general can be attributed to two main factors. First, a moderate fraction of the sample stars evolve off the main-sequence very early in the simulation due to their high mass: $\backsim$$1300$ of the stars reached the end of their main-sequence lifetime before $1$ Gyr. Second, there is a larger population of less massive stars which have small cross-sectional areas, leading to a longer collision timescales.

\subsection{Mass Demographics}

\begin{figure*}[htb!]
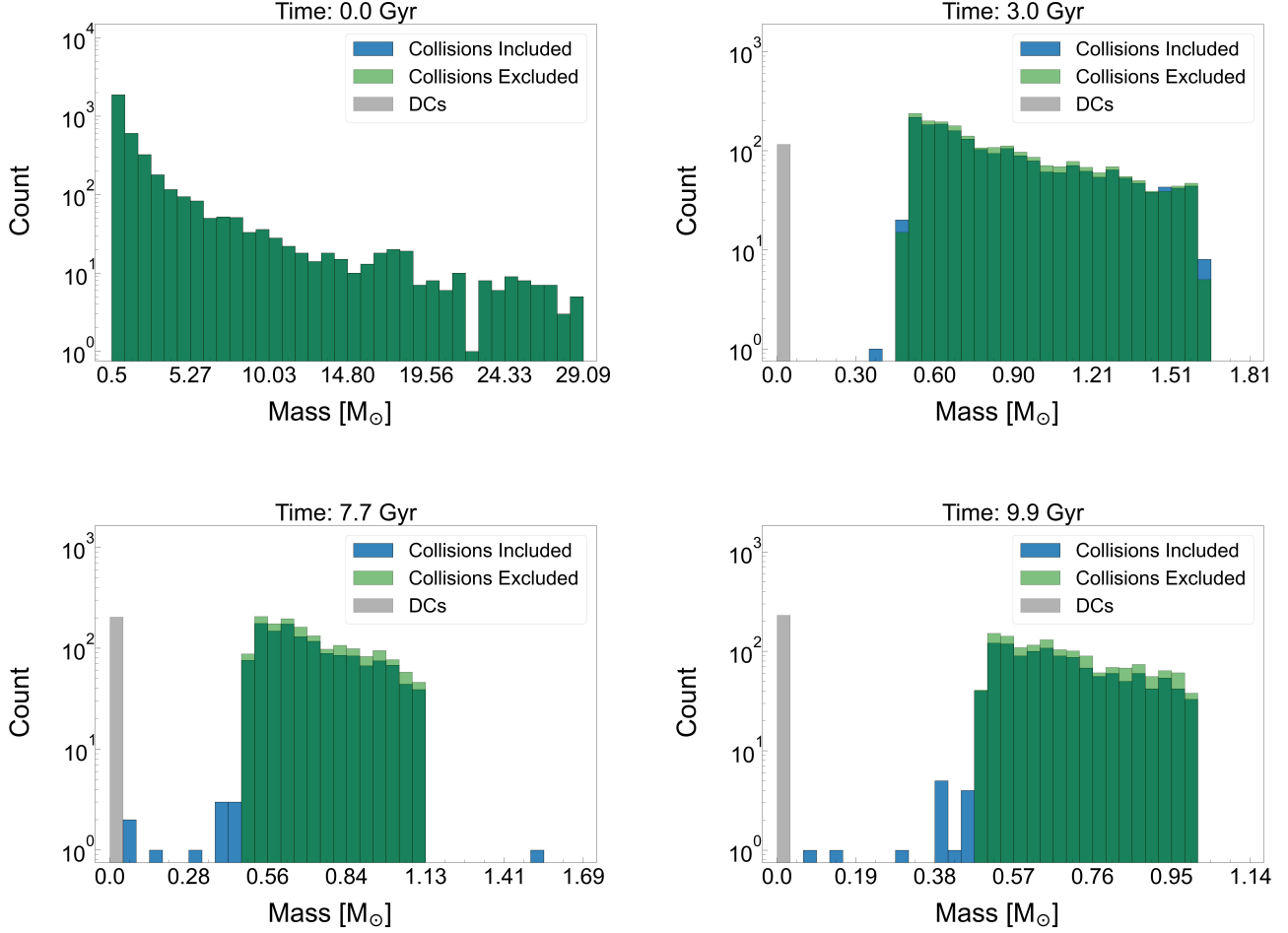

\begin{center}

\resizebox{\textwidth}{!}{%
\begin{tikzpicture}[ultra thick]
    \node[anchor=south west, inner sep=0] (img) at (0,0) {
        \plottwo{PaperPlots/mass_evolution_MS_log_frame_0.pdf}{PaperPlots/mass_evolution_MS_log_frame_1.pdf}
      };
      \begin{scope}[x=1cm,y=1cm]


  \end{scope}
\end{tikzpicture}%
}

\vspace{0.5cm}

\resizebox{\textwidth}{!}{%
\begin{tikzpicture}[ultra thick]
    \node[anchor=south west, inner sep=0] (img) at (0,0) {
        \plottwo{PaperPlots/mass_evolution_MS_log_frame_2.pdf}{PaperPlots/mass_evolution_MS_log_frame_3.pdf}
      };
      \begin{scope}[x=1cm,y=1cm]


  \end{scope}
  
\end{tikzpicture}%
}



  

\end{center}

\caption{Mass distribution evolution for the IMF model. Here, the non-collisional evolution of the cluster is shown in light green and the overlap with the collisional simulation is shown in dark green. As shown, all high-mass stars almost immediately die out due to their relatively short lifetime. As the simulation continues, the remaining massive stars ($>2$ M$_{\odot}$) either die out or experience stripping collisions that decrease their mass into the more populated region. The left-most bar of the histogram represents stars that experienced DCs. We are also able to see the right-most bars rising for short periods of time, indicating merger products, and a population of stripped stars that exist at masses below the minimum initial mass. This plot is animated in the digital copy of the paper.}
\label{fig:mass_evolution_MS}
\end{figure*}

In Figure \ref{fig:mass_evolution_MS}, we examine the mass demographics of the stars over time. Similar to Figure~\ref{fig:mass_evolution}, we show the mass distribution of our sample population in blue. Here, we over-plot in green the evolution of the initial mass function in the absence of collisions, reflecting only changes due to stars evolving off the main-sequence. As expected, after $3$~Gyr, only the lower mass stars ($\lesssim 2$ M$_{\odot}$) remain. 

Similar to the uniform mass models, we count the number of DCs in the gray bin of each histogram. The majority of stripped stars cannot be discerned in the figure because they lie in the initially populated mass range. However, the stripped stars that lie below our minimum initial mass ($0.5$~M$\odot$) can be seen to the right of the DCs. 
We note, however, that these stripped stars may not look like conventional stars of their corresponding masses, a fact that is not immediately apparent from the mass distributions shown in the figure. Previous studies have shown that stripped stars may appear brighter and bluer than other stars of similar mass \citep[e.g.,][]{Gotberg+23, DuttaKlencki24}.

A large portion of stars do not experience collisions, so there are minimal changes in the overall shape of the mass distribution relative to the non-collisional evolution. Over the course of the simulation, we see a few merger products that are akin to blue stragglers \citep[][]{Sills+97, Sills+01, Lombardi+02}. Due to their higher mass, these stars are short-lived in the context of our simulations. These blue stragglers, as seen in the top-right panel, sit at masses slightly higher than the upper mass limit of the stars given the age of the cluster.

\subsection{Destructive Collision Properties} \label{sec:DC_IMF_discussion}

We find similar results for DCs in the simulation with an IMF compared to the uniform mass models. Most stars experience at least 5 collisions prior to destruction and, on average, experience 9 collisions prior to destruction. Given the similarities of these results to previous simulations, we show the corresponding plots in Appendix~\ref{app:MSResults}.

The DC properties are also similar to those from the uniform mass models. The collisions are high speed (average relative velocity of $5600$ km/s) and very head-on (average normalized periapsis separation of 0.2). However, the DCs occur at slightly slower speeds and smaller periapsis separation than the uniform mass model, likely due to the larger population of low mass stars initially present in the simulation. The low mass stars ($<1$ M$_{\odot}$) require fewer collisions to chip away at their mass. Furthermore, they have lower binding energies, making them easier to disrupt. 

\subsection{Mergers and Tidal Disruption Events}

In this model, $3\%$ of stars experienced a merger, but just over $6\%$ of the stars had a net mass gain by the end of their lifetime. We find that approximately half of the stars that gained mass did so through low-speed, non-merger collisions. These collisions, predicted to have two surviving stars by \verb|collAIder|, resulted in small amounts of mass, typically on the scale of $\sim 0.01$ M$_\odot$, being accreted onto our sample star without disrupting the other collider. Notably, this behavior appears when introducing an IMF to the system as it requires the colliding stars have mass ratio $q \neq 1$.
In Figure~\ref{fig:mass_change_MS}, there are numerous stars that peak slightly above their initial mass ($\Delta M \approx +0.01$ M$_\odot$), shown most clearly in the bottom panel of the plot. 
The largest mass gain throughout the simulation was $2.86$ M$_{\odot}$, caused by a single merger within the first $1$ Myr of the simulation. Similar to the uniform mass case, stars often gain mass only to lose it through later stripping collisions. 

This model had a similar TDE rate to the previous simulations. Of the stars that experienced a TDE, 
about one had lost more than $10\%$ of their initial mass and two had gained mass. One of these stars gained mass through a low-speed, non-merger collision, and the other experienced a merger followed by a large number of hit-and-run collisions, resulting in very little overall mass gain ($\Delta M\approx +0.2$ M$_\odot$). As discussed in Section~\ref{sec:TDE_implications_UMM}, collision-affected stars may have unusual compositions and, in the case of merger products, enhanced magnetic fields, all of which may manifest in the observable signatures of TDEs \citep[e.g.,][]{Schneider+16, Schneider+19, Schneider25, Bradnick+17, Gibson+24, Pacuraru+26, Vynatheya+26}.

\section{Ejected Gas from Stellar Collisions} \label{sec:ejected_gas}
\begin{figure*}[th!]
\centering
\includegraphics[width=0.66\textwidth]{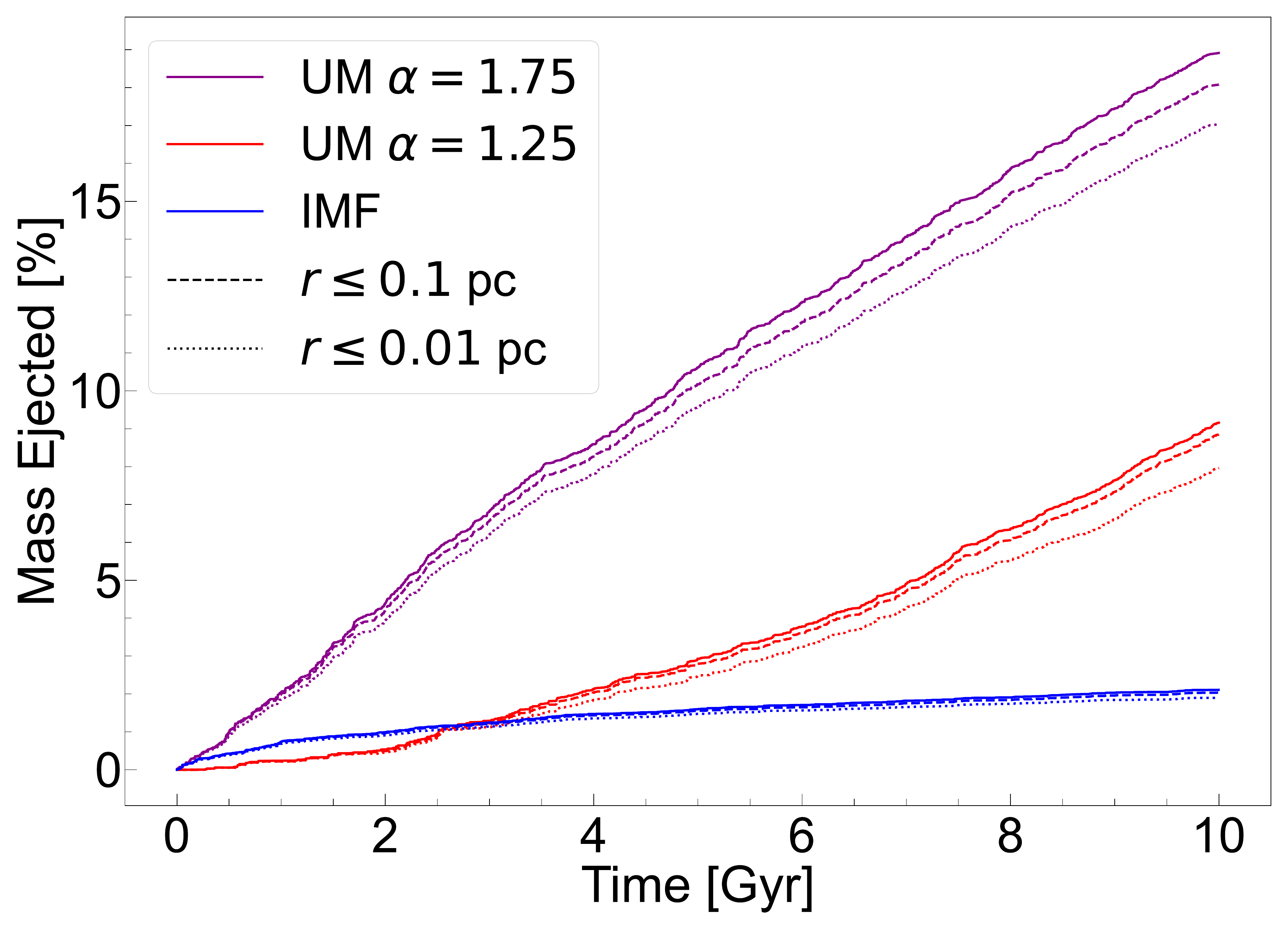}
\caption{Cumulative ejected mass over the course of the simulation for all three cluster models as a percentage of the total stellar mass of the cluster. The uniform mass, $\alpha = 1.75$ model is colored in purple, $\alpha = 1.25$ in red, and the IMF model in blue. For all three models, the dashed and dotted lines denote mass ejected within $0.1$ pc and $0.01$ pc of the SMBH, respectively.}
\label{fig:ejected_gas_vs_time}
\end{figure*}

Lastly, we examine the gas injected into the cluster from mass loss during stellar collisions. In Figure \ref{fig:ejected_gas_vs_time}, we plot the cumulative ejected mass as a percent of the total stellar mass in the cluster for all three previously discussed models. The uniform mass, $\alpha = 1.75$ model produces the largest fraction of ejecta, nearly $20\%$ of the initial stellar mass, owing to this model having the highest number of collisions of the three simulations. The uniform mass, $\alpha = 1.25$ model has the next highest percentage of ejected mass, reaching a total of $9\%$ of the initial stellar mass. The IMF model follows with just over $2\%$ of the stellar mass being injected into the cluster as collision ejecta. Scaled to the size of the physical Galactic center, this corresponds to approximately $760000$ M$_\odot$, $370000$ M$_\odot$, and $250000$ M$_\odot$ respectively. These results trace the overall number of collisions in each simulation. Interestingly, the IMF model has more ejected mass at early times in the simulation compared to the $\alpha = 1.25$ uniform mass model, which overtakes it just after $2$~Gyr. 

For the IMF model, the rate of mass injection slows as the more massive stars in our sample population begin to die off. At very early times ($\lesssim 0.2$~Gyr), both $\alpha = 1.75$ models have a very similar slope, but quickly diverge due to the shorter evolutionary lifetimes of many stars in the IMF simulation; about $1600$ stars have main-sequence lifetimes shorter than 3 Gyr. In the same figure, we also plot the cumulative amount of mass enclosed within $0.1$ and $0.01$ pc from the SMBH. The majority of gas is deposited in the inner region of the cluster, where collisions at high speeds are most likely to occur. In all simulations, over $95\%$ percent of the total mass is ejected within the inner $0.1$~pc of the cluster and over $87\%$ percent within the inner $0.01$~pc.

\begin{figure*}[th!]
\centering
\begin{tikzpicture}
    \node[anchor=south west, inner sep=0] (image) at (0,0) {\includegraphics[width=0.70\textwidth]{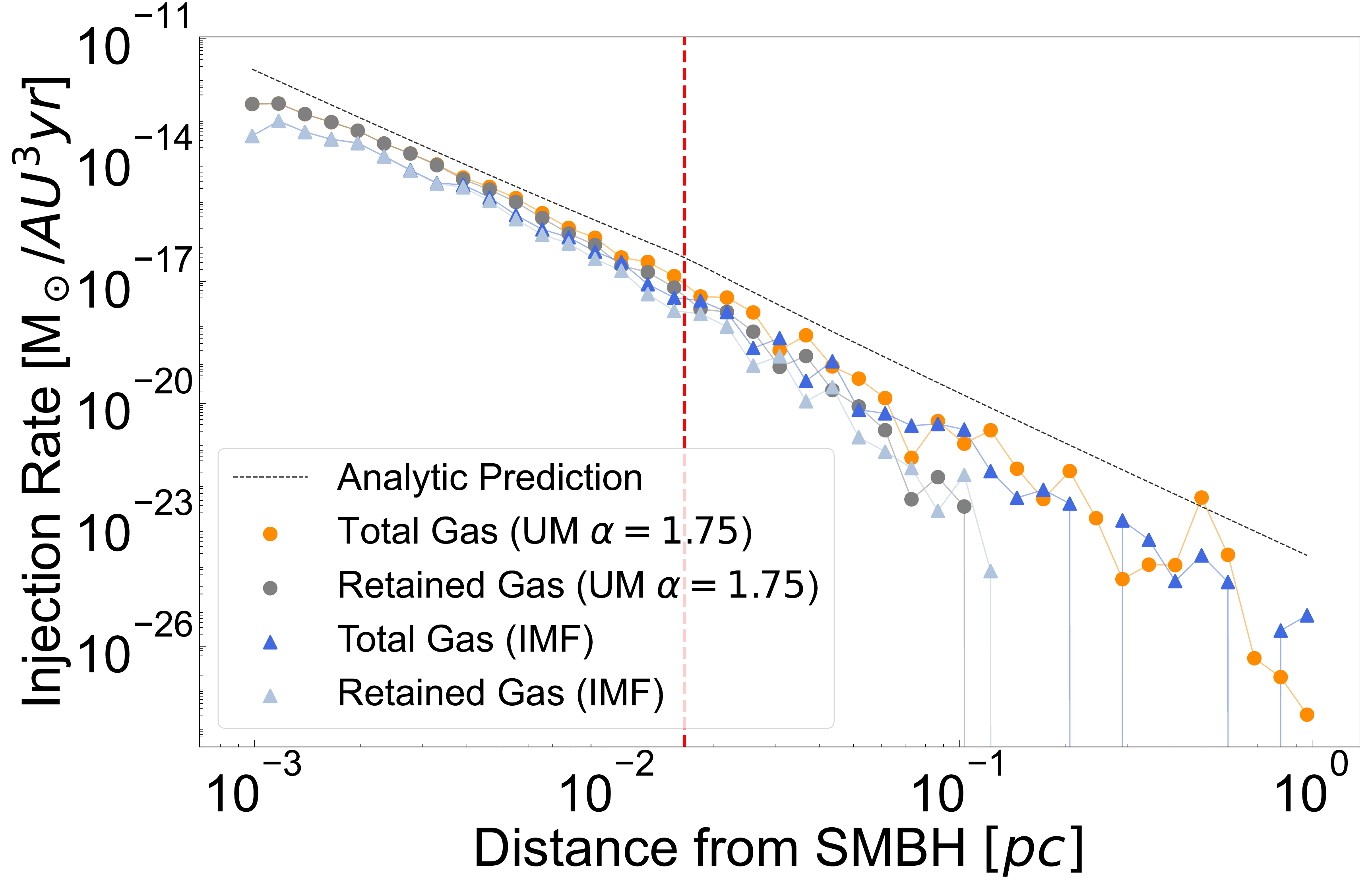}};

    \begin{scope}[x={(image.south east)},y={(image.north west)}]
    
        \draw[->, ultra thick, >=stealth, red] (0.53, 0.8) -- (0.78, 0.8) node[midway, above, font=\Large] {$\lesssim 50\%$ retained};
        
    \end{scope}
\end{tikzpicture}
\caption{Average gas density injection rate for both $\alpha = 1.75$ models. For the uniform mass model, the total injection rate is represented in orange and Equation~\ref{eq:bound_fraction} is applied to determine the retained gas, represented in gray. Similarly, the total injection rate for the IMF model is represented in dark blue and the retained gas in light blue. The analytic prediction is given by Equation~\ref{eq:injection_rate_estimate}. The dashed red line in the middle of the plot indicates the distance at which, on average, $50\%$ of the mass from the collision remains gravitationally bound. For reference, the average density of the interstellar medium is $\backsim$$10^{-18}$ M$_\odot /$AU$^3$.}
\label{fig:injection_density}
\end{figure*}

Additionally, we examine the gas density injection rate of the cluster. In Figure \ref{fig:injection_density}, we estimate the rate as a function of distance from the SMBH for both $\alpha=1.75$ simulations. Here, we scale our results by $1000$ to more accurately represent the Galactic center ($4\times 10^6$ stars). In addition to the overall injection rate, in units of $\mathrm{M}_\odot/\mathrm{AU}^3$ per year, we also plot the fraction of the gas that is retained in the cluster based on the distance from the SMBH at which it is injected. We include an analytic estimate for the the injection rate of the gas, assuming a uniform mass cluster of $1$ M$_\odot$ stars. We express this analytic curve as a function of the collision rate, the number density of the stars, and estimated mass loss fraction per collision. Heuristically, the mass loss fraction can be approximated as the ratio of the kinetic energy of the collision to the binding energies of the stars involved \citep[e.g.,][]{Lai+93,Rose+23,Rose+26,Williams+26}. For Sun-like stars, our analytic estimate can be written as:
\begin{eqnarray} \label{eq:injection_rate_estimate}
     \eta_{gas} &=& t_{coll}^{-1}\times n(a_{\bullet}) \nonumber \\ &\times& \frac{\mu\sigma^2}{GM_*^2/R_* + GM_{\odot}^2/R_{\odot}} \times (M_*+M_{\odot}).
\end{eqnarray}
We plot the analytic prediction in the black dashed line in Figure~\ref{fig:injection_density}. It has a similar slope and shape to our simulation results, though it overestimates the mass injected; the energy ratio is known to overpredict the mass loss from collisions compared to hydrodynamic simulations \citep[e.g.,][]{Lai+93,Rauch99,FreitagBenz,Rose+26}.

The rate of injected gas is higher near the SMBH, where collisions occur more often and tend to be more destructive. We compare our data and analytic curve to an estimate of the retained gas. The fraction of gas from collisions that is bound to the SMBH, and remains in the sphere of influence, is given by
\begin{equation} \label{eq:bound_fraction}
    f_{\rm bound} = \frac{1}{2}\left[1 + \frac{v_{\rm orb}^2 - v_{\rm eject}^2}{2v_{\rm orb}v_{\rm eject}}\right],
\end{equation}
where $v_{\rm orb}$ is the orbital speed of the star during the collision, $v_{\rm eject}$ is ejection speed of unbound material from the star (assumed here to be the escape speed from its surface, $v_{\rm esc} \approx 600$ km/s for a $1$ M$_\odot$ star), and $ 0 \leq f_{\rm bound} \leq 1$ (i.e. if $f_{\rm bound} < 0$, we assume that $0\%$ of the mass is bound and similarly if $f_{\rm bound} > 1$, we assume that $100\%$ of the mass is bound). We can gain physical insight into the system using the velocity dispersion $\sigma$, which is roughly equivalent to Keplerian speed about the SMBH within the sphere of influence at a given position. We estimate $v_{\rm orb}$ as the velocity dispersion $\sigma$ at the distance of the stars from the SMBH when the collision occurs. 

The velocity dispersion becomes equivalent to the escape speed from a Sun-like star at roughly $0.02$ pc, where the bound fraction becomes exactly $50\%$. For our uniform mass model, internal to this distance (indicated with the vertical red line in Figure~\ref{fig:injection_density}) most of the stellar ejecta will remain bound to the SMBH. External to this distance less than half of the ejecta will be bound, and that fraction will decrease with distance from the SMBH. If we assume a higher ejection speed for the stellar material following the collision, the vertical red line will shift closer to the SMBH, and less gas will be retained within the cluster. As most collisions occur and deposit gas near the SMBH (see Figure~\ref{fig:ejected_gas_vs_time}), most of the ejecta remains bound. The bound fraction (Eq.~(\ref{eq:bound_fraction})) assumes isotropic mass loss due to a collision, which becomes less accurate for grazing hit-and-run encounters. Additionally, we assume that the ejecta speed is uniform and fixed for all collisions and ignore factors such as impact speed. However, for the purpose of this study, the bound fraction serves as a useful estimate for the amount of ejected gas that remains bound to the SMBH.

For all models, about $90\%$ of the mass ejected from collisions remains bound to the SMBH. This retention fraction corresponds to $16\%$ of the total initial stellar mass in the cluster for the uniform mass $\alpha = 1.75$ simulation and to slightly less than $2\%$ of the total stellar mass for the simulation with an IMF. 
In the uniform mass model, stellar collisions inject $7.5\times10^{-5}$ M$_\odot$/yr, while they inject $3.2 \times 10^{-5}$ M$_\odot$/yr in the IMF model. The slightly lower rate global injection rate is likely a result of less collisions in the IMF model for the same reasons mentioned previously.

\section{Conclusion} \label{sec:conclusion}

Taking a Monte Carlo approach, we follow a sample of $4000$ stars embedded in the inner pc of the Galactic center and track their evolution due to direct collisions and two-body relaxation. In a departure from previous studies, we utilize \verb|collAIder|,  a ML model trained on SPH simulations of stellar collision, to predict the outcome of stellar collisions \citep{Prieto+26}. This novel approach allows us to accurately predict the outcomes of stellar collisions, up to $\backsim$$15000$ in our simulations, without introducing large computational cost or making simplified assumptions. We run three simulations, varying the slope of the cluster's density profile and the initial mass function of the cluster. We highlight key results in Table~\ref{tab:model_summary}. We find the following:

\begin{enumerate}
    \item \textbf{Stripped stars:} 
    Of the $10^3$ to $10^4$ collisions that occurred in each of our simulations, hit-and-run collisions were by far the most common type. This type of collision results in two surviving stars which may have lost some amount of mass from the impact.
    Across all models, $\gtrsim$$10\%$ of the stars in  the cluster lost at least $10\%$ of their mass due to collisions with other stars. 
    The prevalence of stripped stars has implications for stellar populations, TDE demographics, and the spectral signatures of disrupted stars \citep{Mockler+24}. 
    We expect stripped stars to have a higher metallicity due to ejection of material from the primarily hydrogen outer layers of the star during the impact, shown in detail using SPH simulations by \citet{Gibson+24}. These stars can be found throughout the cluster and even outside the sphere of influence. When compared to other stars of similar mass, these stars may also appear brighter and bluer \citep[e.g.,][]{Gotberg+23, DuttaKlencki24}.

    \item \textbf{Merger products:}
    A few to $5\%$ of the stars in our sample experienced a merger during the simulation.  Mergers are a rarer outcome relative to hit-and-run collisions, and many of the merger products later experience high-speed, stripping impacts, making it difficult to form extremely massive merger products. The most massive stars formed in our simulations through collisions are around $3$~M$_\odot$. It is possible that repeated mergers produce more massive stars in a cluster with $\gtrsim10^6$ stars, but these sequences are too rare to occur in our sample of $4000$. Based on our results, we estimate that $\sim 10^5$ merger products can form within the inner pc of Galactic center. However, only $\sim10^3$ will be the product of a sequence of two or more mergers.
    The merger products may appear younger than the rest of the stellar population, akin to blue stragglers \citep{Sills+97, Sills+01, Lombardi+02,Dray&Tout07, Schneider+14b}.

    \item \textbf{Comparison to previous prescriptions:} We find that we are able to achieve qualitatively similar results 
    compared to previous studies \citep{Rose+23,RoseMockler+25} while using ML to predict the outcomes of collisions. The most destructive collisions occur near the SMBH, while merger products form at distances $\gtrsim0.01$~pc. However, compared to the \citet{Lai+93} fitting formulae, we find a large population of both stripped stars and merger products. Compared to the \citet{Rauch99} prescription, the simulation results in more DCs and fewer mergers.

    \item \textbf{Properties of destructive collisions:} 
    We find that $7$ to $16\%$ of the stars in our simulation are destroyed by a collision. These stars often experience a high number of collisions ($>10$) prior to destruction. These collisions either chipped away at the mass of the star, or they were too grazing to affect the star significantly. The final DCs are often nearly head-on and very high speed ($\backsim$$10 \times$ the escape speed from the surface of the star), collisions which predominantly occur near the SMBH, at distances $<0.01$~pc. 

    \item \textbf{Connections to TDEs:} Collision-affected stars are disrupted by the SMBH at a rate of $\sim 10^{-7}$ per year in our simulations, lower than the overall TDE rate for an SMBH in this mass range by about two orders of magnitude \citep{hannah_counting_2024,chang_rates_2024}. Stripped stars that experience TDEs have been shown to exhibit unique chemical abundances and may serve as ideal candidates for progenitors of observed peculiar TDEs such as ASASSN-14li, PTF15af, and iPTF16fnl \citep{Cenko+16, kochanek_tidal_2016, Blagorod-nova+17, Blagorod-nova+19, Yang17, Gibson+24}. Furthermore, merger products from stellar collisions may have enhanced magnetic fields, which can impact
    the radio emission, x-ray emission, and jet formation of a TDE \citep[e.g.,][]{Schneider+16, Schneider+19, Schneider25, Bradnick+17, Gibson+24, Pacuraru+26, Vynatheya+26}. 
    
    \item{\textbf{Prospects for collision-associated transients:}} DCs have also generated interest as potential transients \citep[e.g.,][]{Ryu+24b,Dessart+24}. The most energetic DCs may produce supernovae-like signals \citep[impact speeds $\sim 10^4$ km/s][]{Balberg+13,BalbergYassur23}. We estimate that DCs with speeds of at least $10000$ km/s occur at rates of $\backsim$$10^{-6}$ per year per galaxy for Milky Way-like Galactic centers. Additionally, luminous red novae (LRNe) have emerged as one of the primary observational manifestations of stellar mergers and common-envelope evolution \citep{Soker&Tylenda03, Kulkarni+07, Tylenda+11, Ivanova+13b, Williams+15, Pastorello+19, Howitt+20}. LRNe have observational timescales of approximately $10$ years \citep{Kaminski+26,Reguitti+26}, and based on the rate of stellar mergers in our simulations it is unlikely that we will observe any merger-induced LRNe in the Milky Way's Galactic center. In the centers of other galaxies, prospects for observing LRNe from stellar collisions are limited by their brightness in a very crowded region of the galaxies, as opposed to their rates \citep{Pastorello+19, Kaminski+26}. 
    However, stellar collisions in general may produce puffy, distended stars similar to the G objects observed in the Milky Way's Galactic center \citep{Antognini15, Witzel+14, Witzel+17, Stephan+16, Stephan+19, Ciurlo+20, Rose+23,Gibson+24}. Assuming stars remain inflated for $\sim10^4$ yr after the collision, we should observe at least one G object from a direct collision in the Galactic center.
    
    \item \textbf{Liberated gas from stellar collisions:}  
    Stellar collisions can inject mass into the cluster, a process that has been considered in the context of Milky Way-like galactic centers by \citet{Murphy+91,DuncanShapiro83,David+87a,David+87b} and \citet{RubinLoeb} and most recently in a cosmological context by \citet{Williams+26}. 
    We find that $95\%$ of the mass is released in the inner $0.1$~pc, where most of the collisions occur.
    Around $85\%$ of the gas remains bound to the SMBH, corresponding to $~10^5$ M$_\odot$ over 10 Gyr when scaled to the size of the Galactic center.
    The ultimate fate of this gas is uncertain. Some of it may be accreted by the SMBH, alongside liberated stellar material from winds of the roughly $30$ Wolf-Rayet (WR) stars in the Galactic center \citep[e.g.,][]{Quataert+99,Cuadra+05,Cuadra+06,Cuadra+08,Ressler+18,Calderon+20,Calderon+25,Lora+21,Balakrishnan+24,Skrabacz+26}. Massive stars release $\sim 10^{-3}$~M$_\odot$ per yr \citep[][]{Quataert+04,Cuadra+08}. While stellar collisions inject less gas into the cluster overall ($\sim10^{-5}$ M$_\odot$ per yr), most of the retained collision ejecta resides within $0.04$~pc of the SMBH, the inner radius of the clockwise disk of young stars in the Galactic center where the WR stars reside \citep[e.g.,][]{Paumard+06,vonFellenberg+22}. Future work will address the fate of the collision ejecta, as well as implications of the introduction of gas for the dynamical evolution of the cluster \citep[see, e.g.,][]{Rozner&RamirezRuiz25}. 
\end{enumerate}

Machine-learning emulators, like \verb|collAIder|, now make it feasible to model these phenomena coupled with the evolution of the Galactic center with a level of physical fidelity previously computationally inaccessible. Our results demonstrate the effectiveness of this tool, and future machine-learning tools, in the study of dynamics of the Galactic center.


\section{Acknowledgments}
This work was supported in part by NSF Grants AST-2108624, AST-2149425, AST-2446392, and AST-2511543 at Northwestern University. We also acknowledge support from the NSF-Simons AI-Institute for the Sky (SkAI) via NSF Grant AST-2421845. We thank E.~Chiang and Z.~Xuan for asking interesting questions about gas liberated through stellar collisions, which inspired part of this work. We are also grateful to E.~Murchikova, E.~Skrabacz, and A.~Ukani for helpful discussions about gas near a SMBH. S.C.R.\ is supported by a Lindheimer Postdoctoral Fellowship. Support for E.G.P.\ was provided by the NSF Graduate Research Fellowship Program under Grant DGE-2234667. F.K.\ acknowledges support from a CIERA Postdoctoral Fellowship.  E.R-R.\ acknowledges support from the Heising-Simons Foundation, and from NSF Grants AST-2206243 and AST-2447606. This research was supported in part through the computational resources and staff contributions provided for the Quest high performance computing facility at Northwestern University, which is jointly supported by the Office of the Provost, the Office for Research, and Northwestern University Information Technology.

\bibliographystyle{aasjournal}
\bibliography{astrophysics_bib}

\appendix

\section{$\alpha = 1.25$ Model Results} 
\label{app:1p25Results}

\begin{figure*}[htb!]
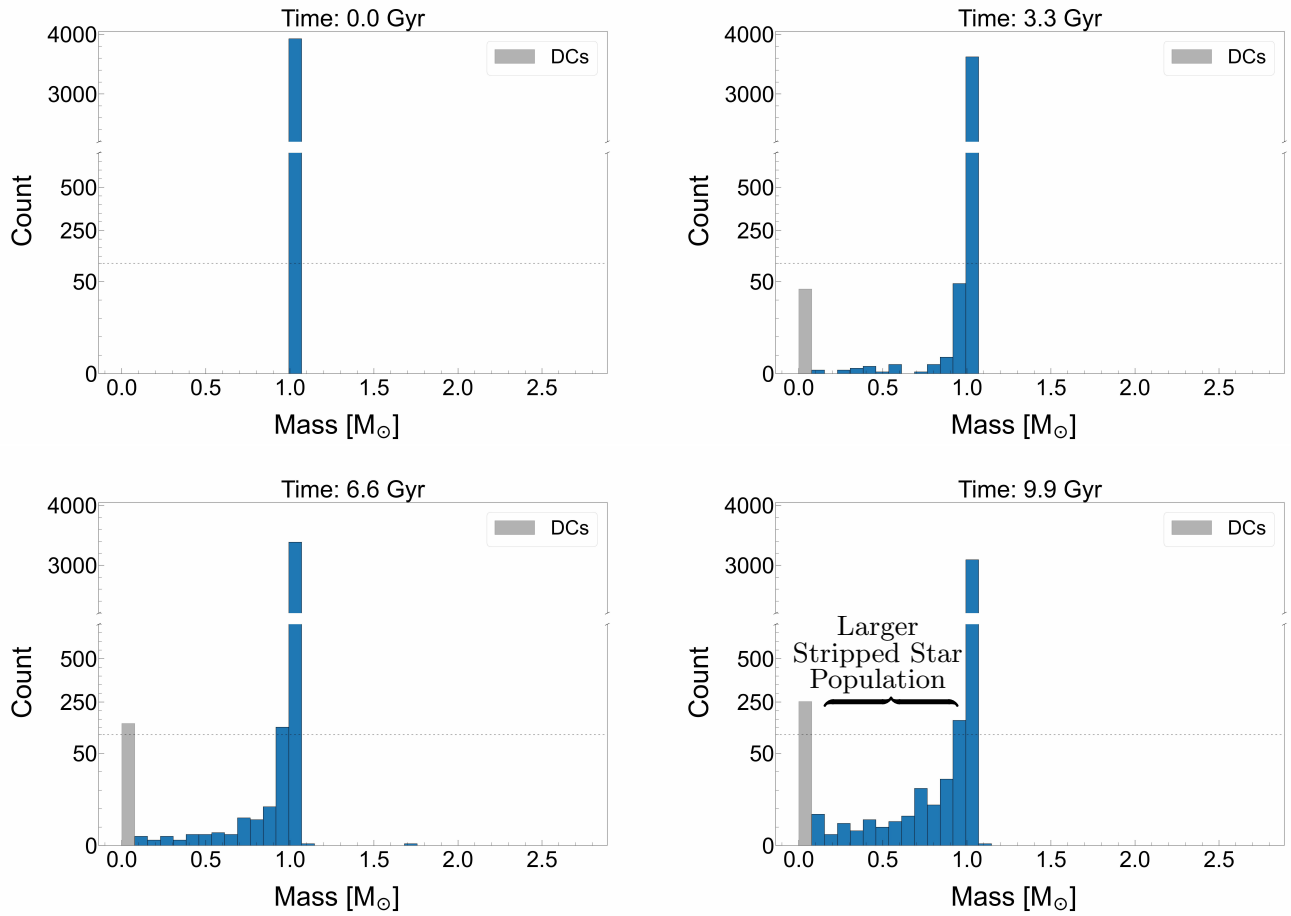

\begin{center}

\resizebox{\textwidth}{!}{%
\begin{tikzpicture}[ultra thick]
    \node[anchor=south west, inner sep=0] (img) at (0,0) {
        \plottwo{PaperPlots/mass_evolution_UM_1p25_frame_0.pdf}{PaperPlots/mass_evolution_UM_1p25_frame_1.pdf}
      };
      \begin{scope}[x=1cm,y=1cm]

  \end{scope}
\end{tikzpicture}%
}


\resizebox{\textwidth}{!}{%
\begin{tikzpicture}[ultra thick]
    \node[anchor=south west, inner sep=0] (img) at (0,0) {
        \plottwo{PaperPlots/mass_evolution_UM_1p25_frame_2.pdf}{PaperPlots/mass_evolution_UM_1p25_frame_3.pdf}
      };
      \begin{scope}[x=1cm,y=1cm]

    \draw [decorate,
    decoration = {calligraphic brace}] (9.5,2.4) --  (11,2.4);
    \node[above] at (10.1,3.0) {Larger};
    \node[above] at (10.1,2.7) {Stripped Star};
    \node[above] at (10.1,2.4) {Population};
  \end{scope}
  
\end{tikzpicture}%
}

\end{center}
\caption{Mass distribution for the $\alpha=1.25$ model over time. Similar to Figure \ref{fig:mass_evolution}, most stars remain at $1$M$_{\odot}$. We see the formation of a stripped star population and a merger product population. We also see significantly less DCs. The stripped star population appears to be slightly larger than that of the $\alpha=1.75$ model. This plot is animated in the digital copy of the paper.}

\label{fig:mass_evolution_1p25}
\end{figure*}

\begin{figure}[H]
\plottwo{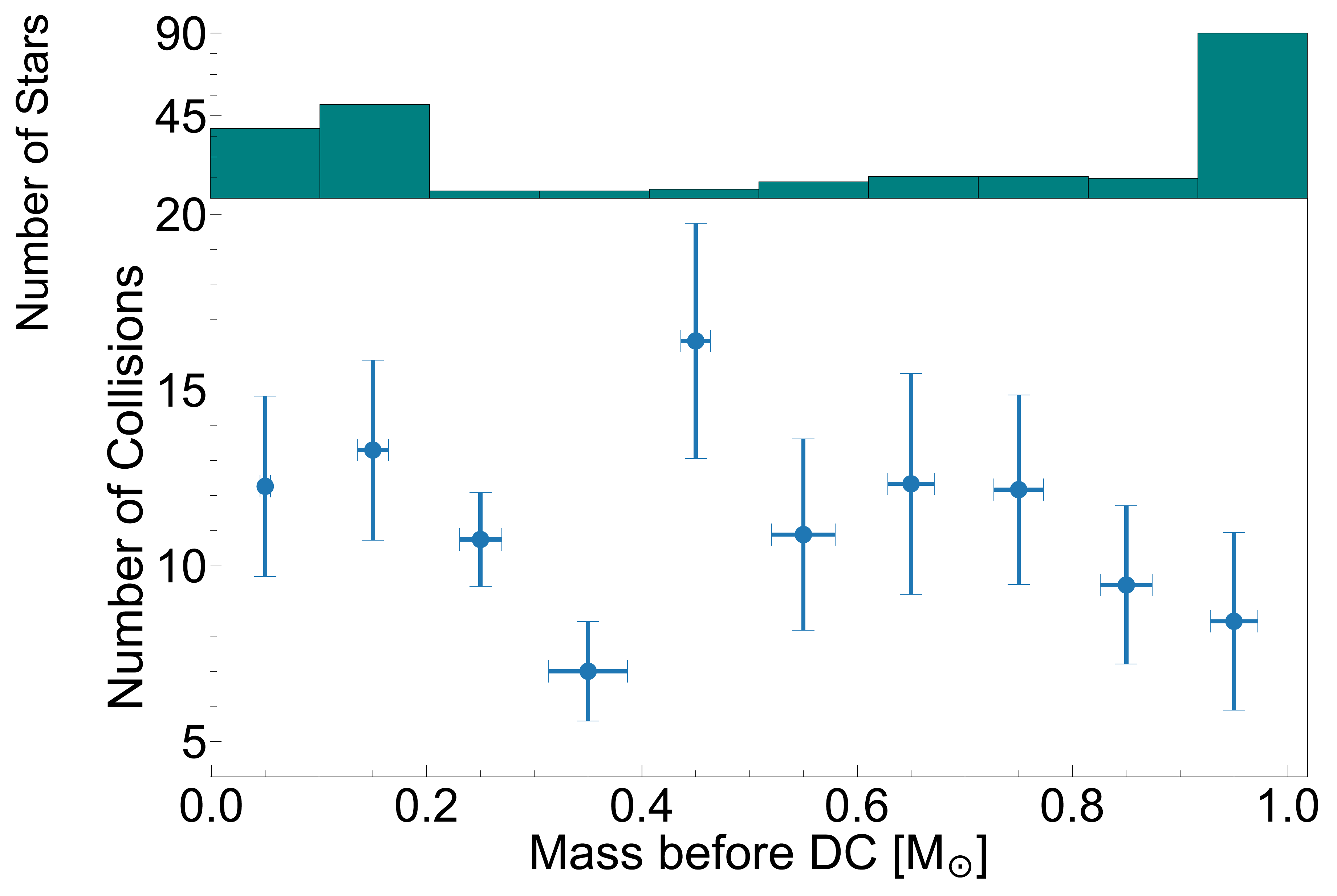}{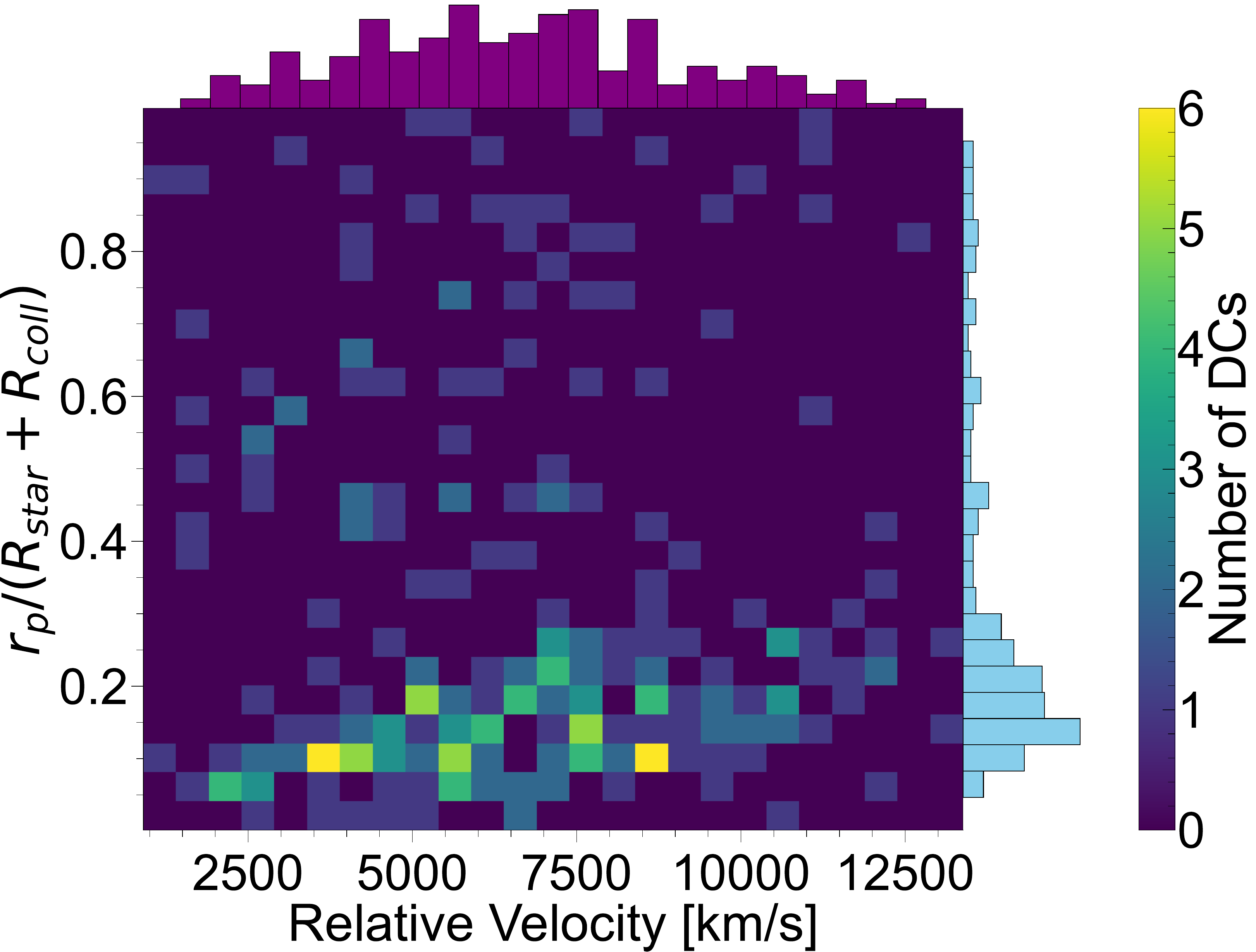}
\caption{\textbf{Mass of stars prior to destruction (Left):} This figure shows the mass of DC stars prior to their final collision for a uniform mass model with an $\alpha=1.25$ stellar density profile. In the top of this plot, we can also see a count of the number of stars in each mass bin. In the bottom of this plot, we see the average number of collisions fro each mass bin. The error in the number of collisions is determined by $\pm\sqrt{\sigma_{count}}$ and the deviation in the mass is determined as $\pm\sigma_{M}$ in mass. \textbf{DC properties (Right):}  In this plot, we see the conditions that lead to DCs for the $\alpha=1.25$ model. Most DCs occur at speeds of at least 2000 km/s and at a normalized impact parameter less than 0.4. We see that the average conditions for an DC is a very high speed collision of approximately $6800$ km/s with a normalized impact parameter of $0.3$.}
\label{fig:DC_Params_1p25}
\end{figure}

In Figure~\ref{fig:mass_evolution_1p25}, we show the mass distribution at different snapshots in time for the $\alpha = 1.25$ uniform mass model. The results are qualitatively similar to the $\alpha=1.75$, but there are fewer merger products and a slightly larger population of stripped stars. 
Figure~\ref{fig:DC_Params_1p25} shows the properties of the DCs, which we describe in Section~\ref{subsec:DC_Properties}. 

\section{IMF Model Results} \label{app:MSResults}

\begin{figure}[H]ß
\plottwo{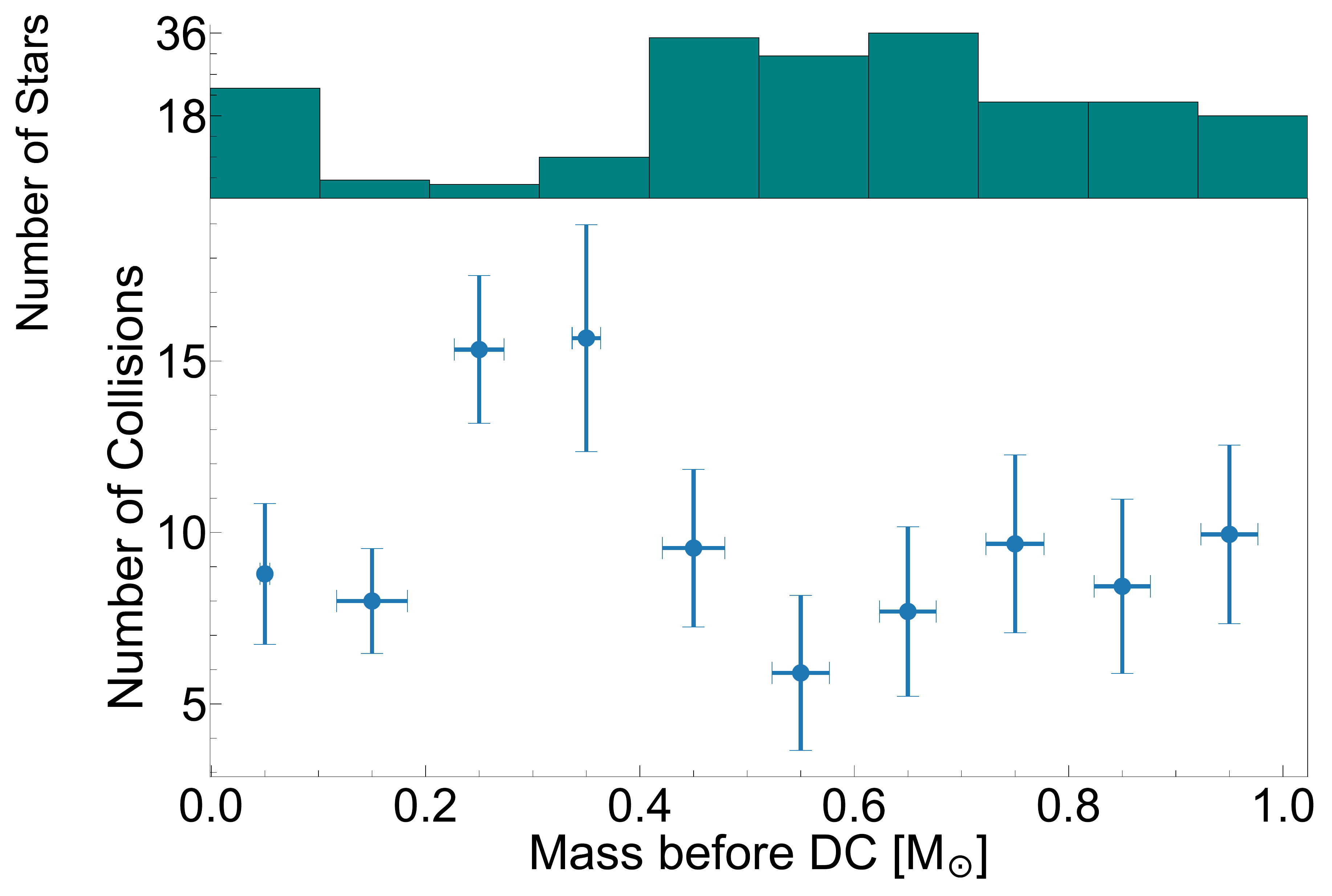}{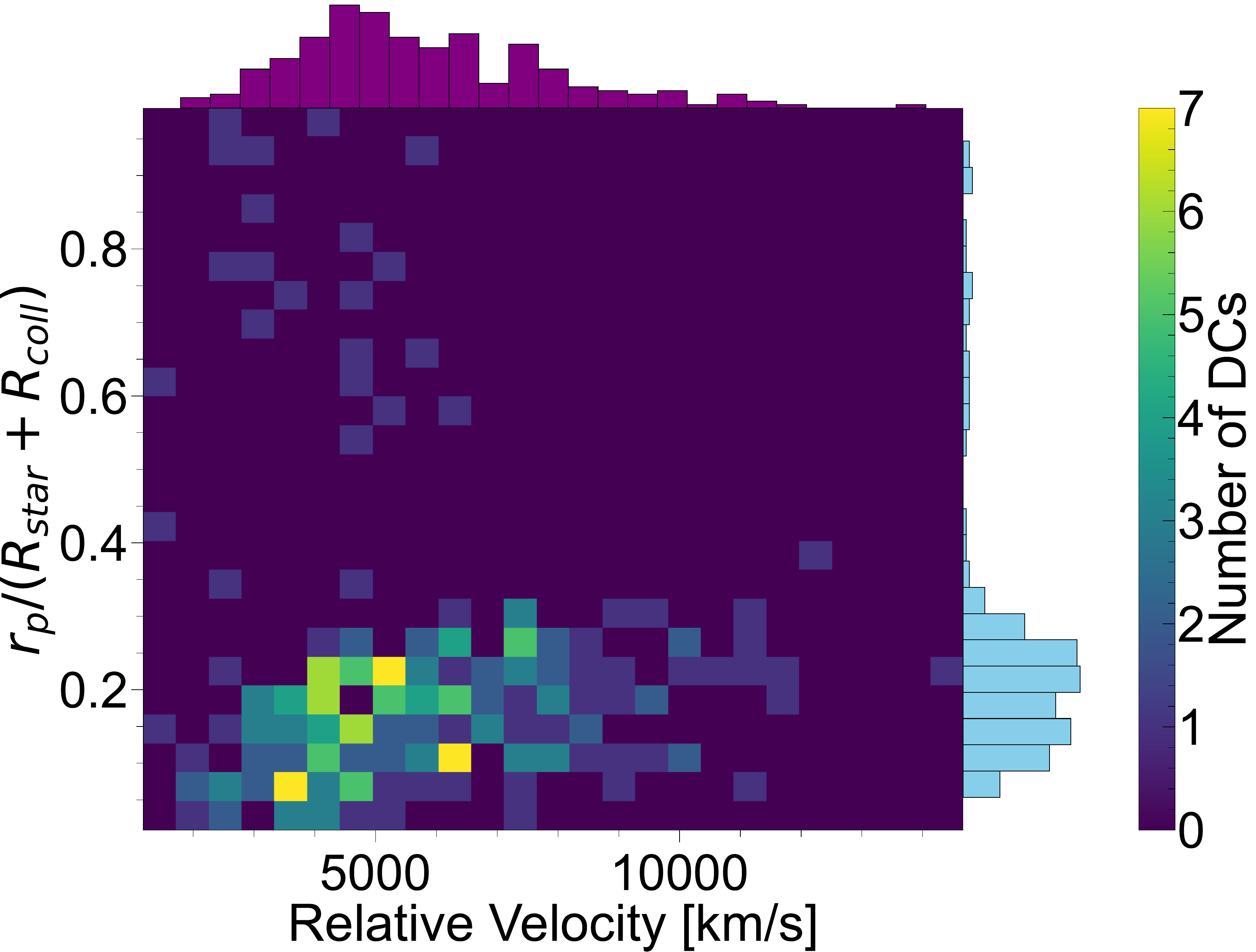}
\caption{\textbf{Mass of stars prior to destruction (Left):} This panel shows the mass of DC stars prior to their final collision for the IMF model. We see a count of the number of stars at each mass. In the bottom of this plot, we see the average number of collisions for each mass bin. The error in the number of collisions is determined by $\pm\sqrt{\sigma_{count}}$ and the deviation in the mass is determined as $\pm\sigma_{M}$ in mass. \textbf{DC properties (Right):} This panel describes the conditions that that lead to DCs for the $\alpha=1.75$ IMF model. As seen in the figure, most DCs occur at speeds of at least 2000 km/s, typically with high directness. We see that the average conditions for an DC is a very high speed collision of approximately $5600$ km/s with a normalized impact parameter of $0.2$.}
\label{fig:DC_Params_MS}
\end{figure}

We show the properties of DCs from the simulation with an IMF in Figure~\ref{fig:DC_Params_MS}, which we also describe in Section~\ref{sec:DC_IMF_discussion}. 
Stars in this model experience an average of 9 collisions before destruction.

\end{document}